\RequirePackage{fix-cm}
\documentclass[twocolumn]{svjour3}          
\smartqed  
\usepackage{graphicx}
\usepackage[utf8]{inputenc}
\usepackage{natbib}
\usepackage[ruled]{algorithm2e}
\usepackage{mathtools}

\DeclareMathOperator*{\argmin}{argmin}
\DeclareMathOperator*{\invlogit}{logit^{-1}}

\graphicspath{{./figures/}}

\begin{document}

\title{Local dimension reduction of summary statistics for likelihood-free inference
}

\titlerunning{Local dimension reduction}        

\author{Jukka Sir\'{e}n         \and
        Samuel Kaski
}

\institute{Jukka Sir\'{e}n \at
              Helsinki Institute for Information Technology HIIT, Department of Computer Science, Aalto University, Espoo, Finland \\
              \email{jukka.2.siren@aalto.fi}           
           \and
           Samuel Kaski \at
              Helsinki Institute for Information Technology HIIT, Department of Computer Science, Aalto University, Espoo, Finland \\
              \email{samuel.kaski@aalto.fi}
}

\date{Received: date / Accepted: date}

\maketitle

\begin{abstract}
Approximate Bayesian computation (ABC) and other likelihood-free inference methods have gained popularity in the last decade, as they allow rigorous statistical inference for complex models without analytically tractable likelihood functions. A key component for accurate inference with ABC is the choice of summary statistics, which summarize the information in the data, but at the same time should be low-dimensional for efficiency. Several  dimension reduction techniques have been introduced to automatically construct informative and low-dimensional summaries from a possibly large pool of candidate summaries. Projection-based methods, which are based on learning simple functional relationships from the summaries to parameters, are widely used and usually perform well, but might fail when the assumptions behind the transformation are not satisfied. We introduce a localization strategy for any projection-based dimension reduction method, in which the transformation is estimated in the neighborhood of the observed data instead of the whole space. Localization strategies have been suggested before, but the performance of the transformed summaries outside the local neighborhood has not been guaranteed. In our localization approach the transformation is validated and optimized over validation datasets, ensuring reliable performance. We demonstrate the improvement in the estimation accuracy for localized versions of linear regression and partial least squares, for three different models of varying complexity.

\end{abstract}

\section{Introduction}
\label{intro}

Approximate Bayesian computation (ABC) and other likelihood-free inference (LFI) methods have gained wide-spread popularity in the last decade \citep{LintusaariGutmannDuttaEtAl2016, SissonFanBeaumont2018}. Beginning with applications in population genetics and evolutionary biology \citep{TavareBaldingGriffithsEtAl1997,PritchardSeielstadPerez-LezaunEtAl1999,BeaumontZhangBalding2002}, the methods have recently expanded to many other fields in science ranging from financial modeling \citep{PetersPanayiSeptier2018} to human-computer interaction \citep{KangasraeaesioeAthukoralaHowesEtAl2017}, and supported by open-source software such as ELFI \citep{LintusaariVuollekoskiKangasraeaesioeEtAl2017}. One of the major contributors to the rise of popularity of the LFI methods is that they allow to connect existing computer simulators to data in a statistically rigorous way. In their simplest form, LFI methods only require the ability to generate pseudo-datasets from a computer simulator and a way to measure the similarity between simulated and observed datasets. 

A key component of any simulation-based likelihood-free inference method is choosing how to measure the similarity between the simulated and observed data sets. The similarity is usually based on low-dimensional summary statistics, which contain most of the information in the data \citep{Prangle2015a}. The low dimensionality is crucial for the good performance of the methods, since they suffer heavily from the curse of dimensionality. For example, under optimal conditions the mean squared error of ABC estimates scales as $O_p(N^{-4/(q+4)})$, where $N$ is the number of samples and $q$ is the dimensionality of the summary statistics \citep{BarberVossWebster2015}.  Design of such summaries is in most cases difficult, which complicates the application of likelihood-free methods to new problems. Some methods bypass the use of summary statistics completely, and work directly on the full data. For example, \citet{GutmannDuttaKaskiEtAl2014} proposed to use classification as a measure for similarity for likelihood-free inference. However, the method is applicable only in situations, where multiple exchangeable samples are available, and hence not generally applicable.
	
Dimension reduction techniques offer a semi-automatic way of producing summary statistics that balances the trade-off between dimensionality and informativeness \citep{BlumNunesPrangleEtAl2012, Prangle2015a}. The most widely used methods are based on a large set of candidate summary statistics. Subset selection methods choose a small subset of the candidate summary statistics, which are the most informative about the parameters \citep{JoyceMarjoram2008, NunesBalding2010}. Projection-based methods construct a functional relationship from summary statistics to the parameters using for example linear regression, and produce new low-dimensional summaries as a combination of the candidate summaries \citep{WegmannLeuenbergerExcoffier2009, AeschbacherBeaumontFutschik2012, FearnheadPrangle2012}.
	
Projection-based dimension reduction techniques face at least two challenges that might compromise their efficiency in producing informative and low-dimensional summary statistics. First, the relationship between the summaries and the target might be more complex than assumed by the dimension reduction method. Usually the relationship is assumed to be linear, which rarely holds globally in the whole parameter space. Second, some of the candidate summaries might be informative only in a subset of the parameter space. This could happen for example in dynamical models, where the amount of data is dependent on the parameter values \citep{SirenLensCousseauEtAl2018}. Consequently, a large dataset allows estimation of more detailed dependencies among the parameters, which would not be possible with a small-sized dataset. Therefore, the optimal summaries for ABC should be different in these regions of the parameter space.
		
The difficulty of applying global projection-based methods can be alleviated by fitting the relationship between summaries and parameters locally around the observed data. This localization may be motivated by the fact that the relationship is usually much simpler when restricted to a smaller region, and hence easier to fit. Also, for estimating the posterior distribution of the observed data, the good performance of the summary statistics is most crucial locally around the data. 

Localization of summary statistics selection has been proposed using at least three different strategies in the literature. In strategy 1, a projection-based transformation is estimated using only simulations that result in datasets close to the empirical data. \citet{AeschbacherBeaumontFutschik2012} suggested performing a pilot ABC analysis using all candidate summaries, and training the boosting with the accepted simulations. Constructing the summary statistics in the neighborhood of the observed data makes it possible to capture the relationship between summaries and parameters more accurately even with a simple model. However, such an approach could perform poorly outside the region of accepted simulations, because after the transformation even simulations outside this region could have similar summaries as the observed data \citep{FearnheadPrangle2012,Prangle2015a}. Additionally, it is not clear how large a set of closest simulations should be used from the pilot analysis. In strategy 2, the prior support is narrowed down to a region of non-negligible posterior density. \citet{FearnheadPrangle2012} suggest performing a pilot ABC run with all candidate summaries, restricting the prior range to a hypercube containing the posterior support and fitting linear regression from the candidates to the parameters. As the prior range is narrowed down, the transformed summaries should behave well in the whole parameter space. A drawback with this approach is that the narrowing down of the prior range might not provide much localization, especially in a high-dimensional setting.
In strategy 3, the localization is achieved by learning a dimension reduction that performs optimally in the neighborhood of the observed data. \citet{NunesBalding2010} introduced a two-stage strategy for selecting a subset of summaries. In the first stage of their method, a number of closest datasets to the empirical one are chosen to be used as validation datasets for the second stage of selecting the optimal subset of summaries. While computationally expensive, this approach has been shown to produce well performing summary statistics \citep{BlumNunesPrangleEtAl2012}, but the validation strategy has not been applied for projection-based methods. 	
	
Here we introduce an algorithm for reliable localization of any projection-based dimension reduction technique. The algorithm combines the localization strategies 1 and 3 described above, and works with any projection based technique. It is  based on first choosing a number of validation datasets, and then optimizing a local projection-based dimension reduction on the validation datasets. The optimization is performed over the size of the neighborhood around the dataset and possible parameters associated with the projection technique, such as the number of components used in partial least squares \citep[PLS,][]{WegmannLeuenbergerExcoffier2009}. By evaluating the performance of the local transformation on the validation datasets globally, the method is able to overcome the issue of poorly performing summaries outside the local neighborhood. We show improvement over global dimension reduction in different models of varying complexity for both linear regression and partial least squares. Compared to the previously published localization approaches, the optimization of the local transformation results in higher accuracy and improved stability of the transformed summary statistics.

\section{Methods}
\label{sec:methods}

Rejection ABC is the simplest algorithm for performing likelihood-free computation \citep{SissonFanBeaumont2018}. It is based on generating $N$ simulated pseudo-datasets from the model $p(D|\theta)$ and comparing those to the observed data. For each simulation $i$ a $d$-dimensional parameter value $\theta_i$ is sampled from the prior distribution and a pseudo-dataset $D_i$ is generated from the model $p(\cdot|\theta)$.  The distance from $D_i$ to $D_{obs}$ is calculated with distance $d(D_i,D_{obs})$, and if $d(D_i,D_{obs})<\epsilon$, for some pre-specified $\epsilon>0$, then simulation $i$ is accepted. The parameters associated with the accepted simulations then constitute an ABC approximation of the posterior distribution $p(\theta|D_{obs})$. As an alternative to specifying a fixed $\epsilon$, many ABC applications instead accept a fixed quantile $\alpha$ of the closest simulations so that the number of samples from the approximate posterior is $\alpha N$.

The distance function $d(\cdot,\cdot)$ is typically defined using  a $q$-dimensional vector of summary statistics $S$, which summarizes the information in the data in a lower-dimensional form \citep{Prangle2015a}. However, in many applications $q$ could be very large, for example hundreds. As discussed above in the introduction, high dimensionality of $S$ provides a challenge for accurate ABC inference. Dimension reduction techniques try to reduce the dimension of $S$ by using a transformation $f(S)$ that produces lower-dimensional summaries that retain most of the information about the model parameters. $f(S)$ may reproduce a small number of elements of $S$, as in subset selection methods that aim to find an informative subset of the summaries \citep{JoyceMarjoram2008, NunesBalding2010}, or $f(\cdot)$ could be a mapping based on estimated relationship from summaries $S$ to parameters $\theta$, as in linear regression \citep{FearnheadPrangle2012} and PLS \citep{WegmannLeuenbergerExcoffier2009} that are the most widely used projection-based dimension reduction methods. In linear regression the parameters are modeled as 
\begin{equation*}
\theta \sim Normal(\mu+S\beta,\Sigma),
\end{equation*}
and the predictions 
\begin{equation}
\hat{\theta} = \hat{\mu}+S \hat{\beta}
\label{eq:linregpred}
\end{equation}
obtained with point-estimate $\hat{\mu}$ and $\hat{\beta}$ are used as summaries. The rationale for using the predictions as summaries is that the posterior mean is an optimal choice for parameter estimation under certain conditions, and the linear predictions give an estimate of the posterior mean.  Dimension reduction techniques require that a sample of $N$ simulations with parameters $\theta$ and summary statistics $S$ is available for estimating the transformation $f(\cdot)$.

\subsection{Local dimension reduction of summaries}

\begin{figure}
	\includegraphics{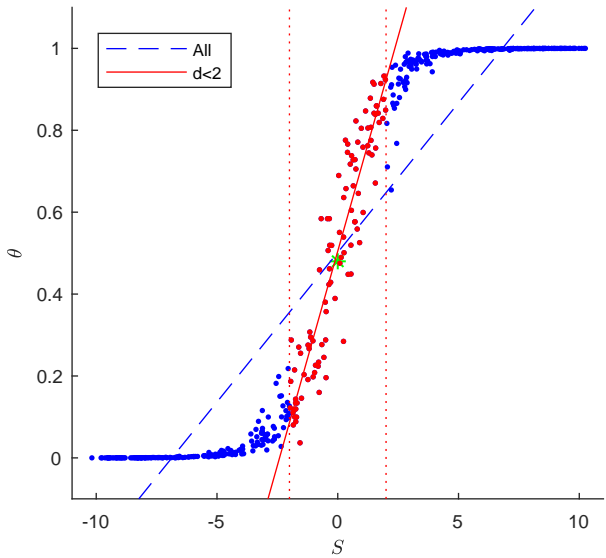}
	\caption{Example of global and local linear regressions. Figure shows the relationship between summaries $S$ and parameters $\theta$ for an artificial example. Each dot represents a simulated data point, and the star represents the observed data. Blue dashed line shows global linear regression line fitted using all simulations, and the red solid line local linear regression using simulations with $d(S,S_{obs})<2$. The dashed vertical lines mark the boundaries of the local neighborhood around the observed data.}
	\label{fig:local}
\end{figure}

Projection based methods of dimension reduction usually aim to find a global mapping from summaries to the model parameters, which does not lose information in the summaries and yet has a simple form. However, the actual relationship is often very complex and such a simple functional form is not possible to obtain. Localization of the transformation, i.e. estimating the projection in the neighborhood of the observed data, provides a solution to this problem, as the relationship is typically less complex within a smaller region. Fig. \ref{fig:local} demonstrates the benefits of localization for linear regression in a simple example. The local linear regression provides a more accurate description of the relationship between $S$ and $\theta$ around the observed data. While the predictions from local linear regression are far off outside the local neighborhood, it should not affect the accuracy of the ABC inference. 

Localization in the data (or summary statistics) space, as proposed by \citet{AeschbacherBeaumontFutschik2012}, is conceptually simple. Instead of using all $N$ simulations for estimating the projection, the estimation is performed using a quantile $\alpha$ of simulations closest to the observed data $D_{obs}$. General pseudo-code for the localization is presented in Algorithm \ref{al:localtransformation}, and an example localization is shown in Fig. \ref{fig:local}.  The user needs to choose the quantile $\alpha$, the initial transformation $f_1$ that is used for defining the set of closest simulations, and any transformation parameters $\lambda$ associated with the transformation $f_l$. The transformation parameters $\lambda$ are separate from the model parameters $\theta$, and they could be, for example, the number of components for PLS. The quantile $\alpha$ defines the amount of localization that algorithm produces. With a small $\alpha$ the produced transformation is more local and should capture the true relationship more accurately around the observed data, but estimating the relationship might be more difficult due to the small number of simulations. \cite{AeschbacherBeaumontFutschik2012} suggest setting $\alpha=500/N$ as a default choice. The initial transformation $f_1$ could be the identity function, resulting in using the set of all candidate summaries as in \citet{AeschbacherBeaumontFutschik2012}, or a transformation from a projection-based method applied globally to all simulations.

\begin{algorithm}[ht]
	\SetAlgoLined
	\SetAlgoNoEnd\SetAlgoNoLine
	\SetKw{KwTo}{in}\SetKwFor{For}{for}{}{end}
	\caption{LocalProjection}
	\KwIn{Initial transformation $f_1$, transformation parameters $\lambda$, size of the local neighborhood $\alpha$, target summaries $S_{obs}$, simulated summaries $S$ and parameters $\theta$}
	\KwOut{Local transformation $f_l$}
	Calculate distances $d(f_1(S_i),f_1(S_{obs}))$ for all simulations $i$\;
	Select set $I_l$ consisting of $\alpha N$ simulations with the smallest distances\;
	Construct transformation $f_l$ based on simulations in $I_l$ using parameters $\lambda$\;
	\label{al:localtransformation}	
\end{algorithm}

The construction of the local transformation $f_l$ on the last line of the Algorithm \ref{al:localtransformation} depends on the projection method used with the algorithm. For example, with regression it amounts to changing the point estimates in the predictions (\ref{eq:linregpred}) to localized ones $\hat{\mu}(I_l)$ and $\hat{\beta}(I_l)$ that  are obtained using only the closest simulations $I_l$. The local transformation is then  
\begin{equation*}
f_l(S) = \hat{\mu}(I_l) +S \hat{\beta}(I_l).
\end{equation*}

\subsection{Optimized local dimension reduction}

The localization Algorithm \ref{al:localtransformation} has the potential to produce more efficient summary statistics than those obtained from global projection methods, but two issues may  lead to poorly-performing summaries. First, the transformation is constructed in the neighborhood of the observed data and should perform well there, but nothing guarantees that the projected summaries are sensible outside this region \citep{Prangle2015a}. The localized transformation might project candidate summaries that are far outside the neighborhood close to the observed data in the lower dimensional space. Second, the size of the neighborhood $\alpha$ used to train the projection should be set somehow, but its optimal value could be almost anywhere between 0 and 1 depending on the model and simulation setting. A default choice, such as $\alpha=500/N$ suggested by \cite{AeschbacherBeaumontFutschik2012}, may work reasonably well in many cases, but might provide sub-optimal and even unreliable results in others. 

We use an optimization strategy similar to the first step proposed by \citet{NunesBalding2010} for the localization, but instead of choosing the best subset of candidate summaries as they suggested, the optimization targets transformation parameters $\lambda$ of Algorithm \ref{al:localtransformation} that include also the size of the local neighborhood $\alpha$. The optimization is based on $N_{valid}$ validation datasets that are chosen as the closest to the observed data after transformation $f_v$. At each step of the optimization a local transformation is constructed for each validation dataset using Algorithm \ref{al:localtransformation}. The performance of the parameters $\lambda$ is evaluated by measuring the accuracy of the posteriors obtained for the validation datasets. The parameters $\hat{\lambda}$ that produce the most accurate posteriors for the validation datasets are then chosen, and the final local transformation targeting the observed data is constructed with Algorithm \ref{al:localtransformation} using $\hat{\lambda}$. While the Algorithm \ref{al:optlocaltransformation} is designed for optimizing the localization, it could also be used without any localization to optimize parameters of a global transformation.

We measure the accuracy of a posterior sample using root mean squared error (RMSE). For posterior sample $\theta_j(I)$ of parameter component $j$ with true value $\theta_{obs,j}$, the RMSE is computed as
\begin{equation}
RMSE\left(\theta_j(I),\theta_{obs,j}\right) = \sqrt{\frac{1}{|I|} \sum_{i \in I} \left(\theta_ {i,j} - \theta_{obs,j} \right)^2}.
\end{equation}
For evaluating the posterior approximation in the whole parameter space, we used summed RMSE, 
\begin{equation}
SRMSE\left(\theta(I),\theta_{obs}\right) = \sum_{j=1}^d RMSE(\theta(I)_j,\theta_{obs,j}).
\label{eq:SRMSE}
\end{equation}
We used average $SRMSE$ over validation datasets as the target for the minimization, although other choices such as maximum over validation datasets could be used as well. Algorithm \ref{al:optlocaltransformation} shows pseudo-code for the optimized local dimension reduction using exhaustive search over a set $\varLambda$ of candidate values for the transformation parameters $\lambda$. An exhaustive search over a grid of candidate values works reasonably well, if the dimensionality of $\lambda$ is small and the grid does not need to be dense. In the general case, the grid search could be substituted for a more efficient optimization algorithm.

The transformation parameters $\lambda$ in Algorithm \ref{al:optlocaltransformation} are not limited to the parameters of the local transformation $f_l$ and the size of the neighborhood $\alpha$. The $\lambda$ could include also parameters corresponding to the initial transformation $f_1$ that is used for localization, as the localization is done separately for each value of $\lambda$ and hence it is possible to optimize $f_1$. The other initial transformation $f_v$ is used before optimization, and therefore, it cannot be optimized within Algorithm \ref{al:optlocaltransformation}.

\begin{algorithm}[ht]
	\SetAlgoLined
	\SetAlgoNoEnd\SetAlgoNoLine
	\SetKw{KwTo}{in}\SetKwFor{For}{for}{}{end}
	\caption{LocalProjectionOptimized}
	\KwIn{Initial transformations $f_v$ and $f_1$, candidate transformation parameters $\varLambda$, number of validation datasets $N_{valid}$, number of samples to approximate posterior $N_{post}$, target summaries $S_{obs}$, simulated summaries $S$ and parameters $\theta$}
	\KwOut{Local transformation $f_l$}
	Calculate distances $d(f_v(S_i),f_v(S_{obs}))$ for all simulations $i$\;
	Select set $I_{valid}$ consisting of $N_{valid}$ simulations with smallest distances \;
	
	\For{$\lambda \in \varLambda$}{
		\For{$i \in I_{valid}$}{
			Construct local transformation $f_{i,\lambda}$ with transformation parameters $\lambda$ targeting dataset $i$ using Algorithm \ref{al:localtransformation} with $f_1$ as the initial transformation.\;
			Calculate distances $d(f_{i,\lambda}(S_{i^*}),f_{i,\lambda}(S_{i}))$ for all simulations $i^* \neq i$.\;
			Select set $I_{post}$ consisting of $N_{post}$ simulations with the smallest distance.\;
			Compute $SRMSE_{i,\lambda} = SRMSE\left(\theta(I_{post}),\theta_{i}\right)$ with equation \ref{eq:SRMSE}\;	
		}
		Compute $SRMSE_{\lambda} = \sum_{i \in I_{valid}} SRMSE_{i,\lambda}$ as a measure of fit for parameter $\lambda$ \;
	}
	$\hat{\lambda} = \argmin_{\lambda \in \varLambda} SRMSE_{\lambda}$ \;
	Construct local transformation $f_l$ with transformation parameters $\hat{\lambda}$ targeting observed dataset using Algorithm \ref{al:localtransformation} with $f_1$ as the initial transformation\;
	\label{al:optlocaltransformation}
	
\end{algorithm}

\section{Example cases}

In this section we apply the developed methods for analyzing simulated datasets under four different models. We compared seven different dimension reduction techniques: linear regression (Reg), local linear regression (localReg), optimized local linear regression (localRegopt), partial least squares (PLS), optimized partial least squares (PLSopt), local partial least squares (localPLS) and optimized local partial least squares (localPLSopt). The aim of the comparison was to study how much localization improves widely used projection-based dimension reduction techniques on different models, and to study the effect of the proposed optimization method on the local dimension reduction techniques. We implemented the dimension reduction methods and performed the example analyses in Matlab\footnote{Code for running the methods and experiments is available at {https://github.com/jpsiren/Local-dimreduc}}.

\subsection{Setting for the examples}

In all cases we normalized the candidate summaries before applying the dimension reductions. We first applied to each non-negative candidate summary a square root transformation, which stabilizes their variance and should make their distributions closer to normal distributions. The partial least squares method used in this work assumes that the candidates have a normal distribution, and it is common to apply a power transformation to the candidates before PLS \citep{WegmannLeuenbergerExcoffier2009}. For linear regression the normality of the predictors is not necessary, but the square root transformation might improve its performance, and allows more direct comparison to partial least squares. After this we standardized each candidate summary to have zero mean and unit variance. In all applications of the Algorithm \ref{al:optlocaltransformation} we set the number of validation datasets as $N_{valid}=20$ and number of posterior samples used within the algorithm as $N_{post}=200$. The posterior distributions were approximated using 100 closest simulations in each case.

In Reg we modeled the model parameters $\theta$ with linear regression using the set of all candidate summaries $S$ as covariates, and used the predictions $\hat{\theta} = S \hat{\beta}$ obtained with ordinary least squares as transformed summaries. In localReg we used similar linear regression, but localized the transformation with Algorithm \ref{al:localtransformation} using the size of the local neighborhood $\alpha=500/N$ following \citet{AeschbacherBeaumontFutschik2012}. In localRegopt we localized linear regression with Algorithm \ref{al:optlocaltransformation} optimizing $\alpha$. As candidate values for $\alpha$ we used $\log_{10}(\alpha_c) = (-1.5,\allowbreak -1.35,\allowbreak -1.2,\allowbreak -1.05,\allowbreak -0.9,\allowbreak -0.75,\allowbreak -0.6,\allowbreak -0.45,\allowbreak -0.3,\allowbreak -0.15)$. The number of candidate values was kept low for computational efficiency. For selecting the validation datasets and localization, we used global linear regression as the initial transformation.

In PLS we fitted partial least squares from $S$ to $\theta$ using the \texttt{plsregress} function implemented in Matlab, and used the transformed PLS components as summaries. We chose the number of components based on  mean squared errors estimated with 10-fold cross-validation. We cut the number of components at the point where inclusion of the next-largest component decreased MSE less than 1 \% of the total variation in $\theta$, but still using at most 15 components. In PLSopt we fitted global PLS similarly as in PLS, but optimized the number of components with Algorithm \ref{al:optlocaltransformation}. In localPLS we performed localized PLS with Algorithm \ref{al:localtransformation} using $\alpha=500/N$ and chose the numbers of components
for both $f_1$ and $f_l$ similarly as in global PLS. In localPLSopt we performed localized PLS with Algorithm \ref{al:optlocaltransformation}, optimizing $\alpha$ among $\alpha_c$, number of components of the local PLS transformation $f_l$, and number of components of the initial PLS transformation $f_1$ used for localization. In all PLS transformations we set the maximum number of components to 15. For selecting the validation datasets, we used the PLS transformation with 15 components as the initial transformation $f_v$.

\subsection{Ricker map}

Ricker map is an ecological model describing the dynamics of a population over time. The model has a relatively simple form, yet it produces highly complex dynamics with nearly chaotic behavior. Inference for such models is difficult with likelihood-based approaches, but ABC and other likelihood-free methods have been proposed as alternatives \citep{Wood2010,FearnheadPrangle2012}. The population size $N_t$ changes over one time step according to
\begin{equation}
N_{t+1} = r N_t exp(-N_t + e_t),
\end{equation}
where $e_t$ are independent noise terms with normal distribution N($0$,$\sigma_e^2$) and $r$ is the intrinsic growth term. Observations $y_t$ from the model at time $t$ are assumed to follow Poisson($\phi N_t$) distribution. The parameters of interest are $\theta = (\log(r),\sigma_e,\phi)$.

In our study we followed \citet{FearnheadPrangle2012} and simulated the Ricker map for 100 time steps from initial state $N_0=1$ with data from the last 50 time steps. We created 100 test data sets using $\log(r)=3.8$, $\phi=10$ and $\log(\sigma_e)$ values from uniform grid between $\log(0.1)$ and $0$. We used independent uniform priors on $\log(r)$, $\log(\sigma_e)$ and $\phi$ with ranges $(0,10)$, $(\log(0.1),0)$ and $(0,100)$, respectively. For the analyses we simulated a total of 1,000,000 datasets with parameter values sampled from the prior distribution. As candidate summaries we used autocovariances and autocorrelations up to lag 5 for $y$, mean and variance of $y$, $\sum_t I(y_t=k)$ for $k=0,...,4$, $\log(\sum_t y_t^i)$ for $i=2,...,6$, logarithms of the mean and variance of $y$, time-ordered observations and magnitude-ordered observations. In total, we had 124 candidate summaries. 

\begin{figure}
	\includegraphics{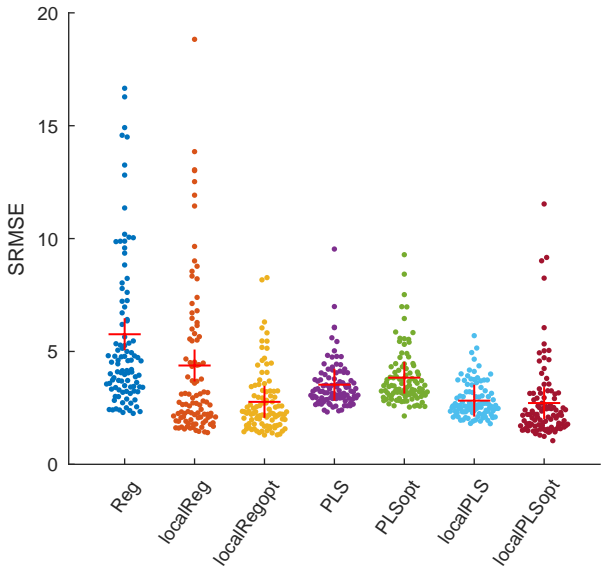}
	\caption{Accuracy of different dimension reduction techniques with the Ricker model evaluated over simulated test datasets. The plot shows the average SRMSE for the 100 test datesets. 'Reg' and 'PLS' refer to global regression and PLS transformations for the parameters. 'PLSopt' refers to PLS with the number of components optimized using validation datasets. 'localReg' and 'localPLS' refer to local versions of the regression and PLS transformation with Algorithm \ref{al:localtransformation}, respectively. 'localRegopt' and 'localPLSopt' refer to optimized local versions of the regression and PLS transformation with Algorithm \ref{al:optlocaltransformation}, respectively.}
	\label{fig:rickerRMSEav}
\end{figure}

With the Ricker model, localization helped to achieve higher accuracy both with linear regression and PLS, compared to the global versions of the transformations for all parameters (Fig. \ref{fig:rickerRMSEav}). The optimized local transformations produced on average higher accuracy than the regular local transformations, but there was some variation for different parameters (Fig. S1 in the Supplementary material). The size of the neighborhood $\alpha$ was on average almost three times smaller with localRegopt compared to localPLSopt (Table S1 in the Supplementary material). The number of PLS components used was highest with the global PLS methods and lowest with localPLSopt (Table S2 in the Supplementary material). 

The long tails of the SRMSE values shown in Fig. \ref{fig:rickerRMSEav}, especially with the regression methods, indicate that the ABC sometimes failed to produce accurate results. We were not able to find any single cause for these failures, but many of the cases seemed to be related to the near-chaotic behavior of the Ricker map. \citet{FearnheadPrangle2012} noted that regression fitted poorly with datasets mostly consisting of 0s and removed datasets with more than 44 0s before the analysis. In our experiment, we kept all datasets and around one quarter of the simulated datasets had only 0, possibly reducing the accuracy of the methods.

\subsection{Individual-based model of bird population dynamics}

We analyzed the individual-based model (IBM) developed by \citet{SirenLensCousseauEtAl2018} for understanding the population dynamics of White-starred robin in Taita Hills forest network in Kenya. The posterior distribution of the model parameters as well as predictions of population state were estimated with an ABC approach in the paper for capture-recapture and genetic data spanning 13 years. We give here a brief overview of the model, but for detailed description see \citet{SirenLensCousseauEtAl2018}. 

The model is spatially structured with 14 habitat patches and surrounded by matrix unsuitable for the species. Each patch with size $A$ is divided to Poisson($qA$) number of territories that can be occupied by a pair (a male and a female). Mating happens once a year in territories occupied by a pair and produces Binomial($2,p^J$) juveniles. After fledgling phase, the juveniles emigrate with probability $\invlogit(\nu_j+\nu_a A)$ and immigrate to another patch $i$ with probability proportional to $e^{-\alpha_I d_i}$, where $d_i$ is the distance to patch $i$. The juveniles become adults after two years and may occupy free territories. Floaters (adults not occupying a territory) emigrate to another patch with daily probability $\invlogit(\nu_f+\nu_a A)$ and have same immigration probabilities as the juveniles. The mortality of individuals is modeled on daily basis with probability $\invlogit(\zeta_d + \zeta_S I(female))$, where $I(female)$ is indicator for females. Each individual carries a genotype in a number of diploid microsatellite loci, and the genotypes follow Mendelian laws with step-wise mutation occurring with probability $\mu$. Observations are made during mist-netting sessions with each individual in the patch having probability $\invlogit(\eta_1+\eta_2 L + \eta_3 A + \eta_4 I(floater))$ of being observed, where $L$ is the sampling intensity and $I(floater)$ is an indicator for floaters. The observation includes the identity of the bird and whether it is a juvenile or an adult, and for a predefined proportion of individuals also genotype and sex. The parameter vector of interest is $\theta = (\log(q),\allowbreak \zeta_d,\allowbreak \zeta_s,\allowbreak \nu_f,\allowbreak \log(-\nu_a),\allowbreak \log(\alpha),\allowbreak \mathrm{logit}(p^J),\allowbreak \nu_j,\allowbreak  \log(\mu),\allowbreak \eta_1,\allowbreak \log(\eta_2),\allowbreak \log(-\eta_3),\allowbreak \log(\eta_4))$.

We used the same set of simulations analyzed in \citet{SirenLensCousseauEtAl2018} in scenario R+G(5,0.2) (5 loci and 20 \% of individuals genotyped) corresponding to the full observed data. This included a total of 100,000 simulations with parameter values drawn from uniform prior distributions and evaluated at 344 candidate summary statistics, and additional 100 test datasets that were simulated with parameters set to produce datasets similar to the observed data. Due to the formulation of the model, the number of observations in the dataset varies strongly with the parameter values, and almost 75 \% (74,830) of all simulations did not produce stable population and hence the datasets contained no observations. Therefore, we used two different sets of simulations to analyze the 100 test datasets: all 100,000 simulations and those 25,170 simulations with observations.

\begin{figure*}
	\includegraphics[width=0.75\textwidth]{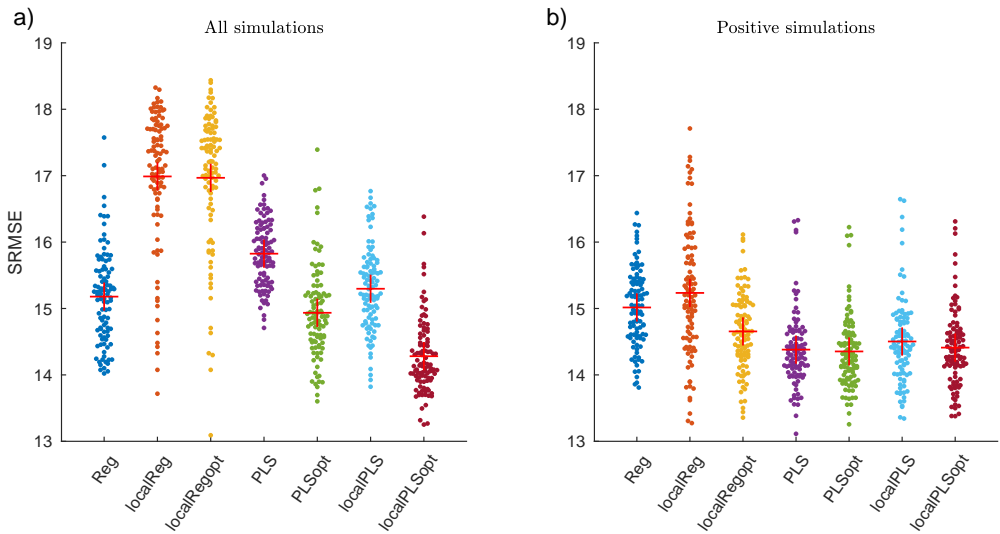}
	\caption{Accuracy of different dimension reduction techniques with the bird population dynamics IBM evaluated over simulated test datasets. The plots shows the average SRMSE for the 100 test datesets using all simulations (a or using only those 25,170 simulations with observations (b). 'Reg' and 'PLS' refer to global regression and PLS transformations for the parameters. 'PLSopt' refers to PLS with the number of components optimized using validation datasets.'localReg' and 'localPLS' refer to local versions of the regression and PLS transformation with Algorithm \ref{al:localtransformation}, respectively. 'localRegopt' and 'localPLSopt' refer to optimized local versions of the regression and PLS transformation with Algorithm \ref{al:optlocaltransformation}, respectively.}
	\label{fig:WRRMSEsum}
\end{figure*}

Using all the 100,000 simulations, localization of the transformations with optimization lead to improved accuracy with PLS, but when restricted to only using the 25,170 simulations with observations the accuracy was similar with the global and local versions of PLS (Fig. \ref{fig:WRRMSEsum}). This was probably due to the low number of simulations, which made it difficult to robustly estimate the relationship from the high-dimensional set of candidate summaries to the parameters.  The performance of the four PLS transformations using only simulations with observations was also similar with optimized local PLS using all simulations.  

For linear regression the results were somewhat mixed. With all simulations localization resulted in significantly higher SRMSE than the global regression, but using only simulations with observations optimized local regression helped to improve accuracy over global version. The difference using localization with the two sets of simulations was that with all simulations the initial regression transformation used for the localization worked poorly. Only a few thousand of the closest simulations contained observations with the number varying with the test dataset, while other simulations with observations had highest distance to the observed data. This did not cause problems with the global regression, because the closest 100 simulations that were used to approximate the posterior were similar to the test data, but made it almost impossible to use a localized version of the regression. 

The non-optimized localization provided slight improvement over global transformation only for PLS with all simulations, while for other combinations it decreased the accuracy compared to the global transformation (Fig. \ref{fig:WRRMSEsum}). The failure of the regular localization was probably due to the size of the neighborhood (500 samples), which was too small for the high-dimensional problem. RMSE values separately for each parameter showed mostly the same patterns as SMRSE, but there were some differences in the variation between the methods among the parameters (Figs S2 and S3 in the Supplementary material). The number of PLS components was high for all PLS methods and both sets of simulations, with the exception of localPLSopt using all simulations, for which the average number was 1.56 (Table S2 in the Supplementary material). This low value combined with the relatively high average for $\alpha$ (0.156, Table S1 in the Supplementary material) further suggests that the localization in this case mostly resulted in removal of the simulations with no observations. In direct contrast to the results for the Ricker model, localRegopt resulted in significantly higher $\alpha$ values than localPLSopt with only simulations with observation (Table S2 in the Supplementary material).

To confirm that the problems with the local linear regressions were related to the initial transformation, we reran the analyses using identity function as the initial transformation (i.e. using all candidate summaries directly) for regular and optimized local regressions. The alternative initial transformation lead to good performance for both local regressions, and optimized local regression had the highest accuracy over all dimension reduction techniques for both sets of simulations (Fig. S4 in the Supplementary material). The non-optimized local regression improved over the global regression for both sets of simulations with SRMSE values in the middle between those of global and optimized local regression (Fig. S4 in the Supplementary material).

\subsection{$g$-and-$k$-distribution}

$g$-and-$k$-distribution is a flexible univariate distribution that can be used to model  skewed and  heavy-tailed data with a small number of parameters \citep{HaynesMacGillivrayMengersen1997}. The likelihood function of the distribution does not have a closed form, but its quantile function is
\begin{align*}
F^{-1}&(x|A,B,c,g,k) = \\
  &A + B\left(1+c \frac{1-\exp(-gz(x))}{1+\exp(-gz(x))}\right)\left(1+z(x)^2 \right)^k z(x),
\end{align*}
where $z(\cdot)$ is the quantile function of the standard normal distribution. Simulation from the model is straight-forward with the inversion method, making the model ideally suited for ABC, and it has been widely used as a test case for new ABC methods \citep{FearnheadPrangle2012, DrovandiPettittLee2015, Prangle2017,SissonFanBeaumont2018}. Typically the parameters of interest are $\theta = (A,B,g,k)$ with restriction $B>0$ and $k>-1/2$, and with $c$ fixed to value $0.8$.

We analyzed 100 simulated datasets with 10,000 samples each from the $g$-and-$k$-distribution with parameters $\theta=(3,1,2,0.5)$ following the study of \citet{FearnheadPrangle2012}. For the ABC based inference we simulated 800,000 pseudo-datasets with parameters sampled from uniform prior distribution on $(0,10)^4$. We used 200 evenly spaced quantiles as candidate summaries for the dimension reduction algorithms.

Additionally, we tested the effect of dimensionality to the performance of the algorithms as a function of both the number of simulations and number of candidate summaries. We considered subsets of the simulations with 25, 50, 100 or 200 candidate summaries and 25,000, 50,000, 100,000, 200,000, 400,000 or 800,000 simulated pseudo datasets. We ran the algorithms on all test datasets, with each combination of the number of candidate summaries and pseudo datasets.

\begin{figure*}[ht]
	\includegraphics[width=0.75\textwidth]{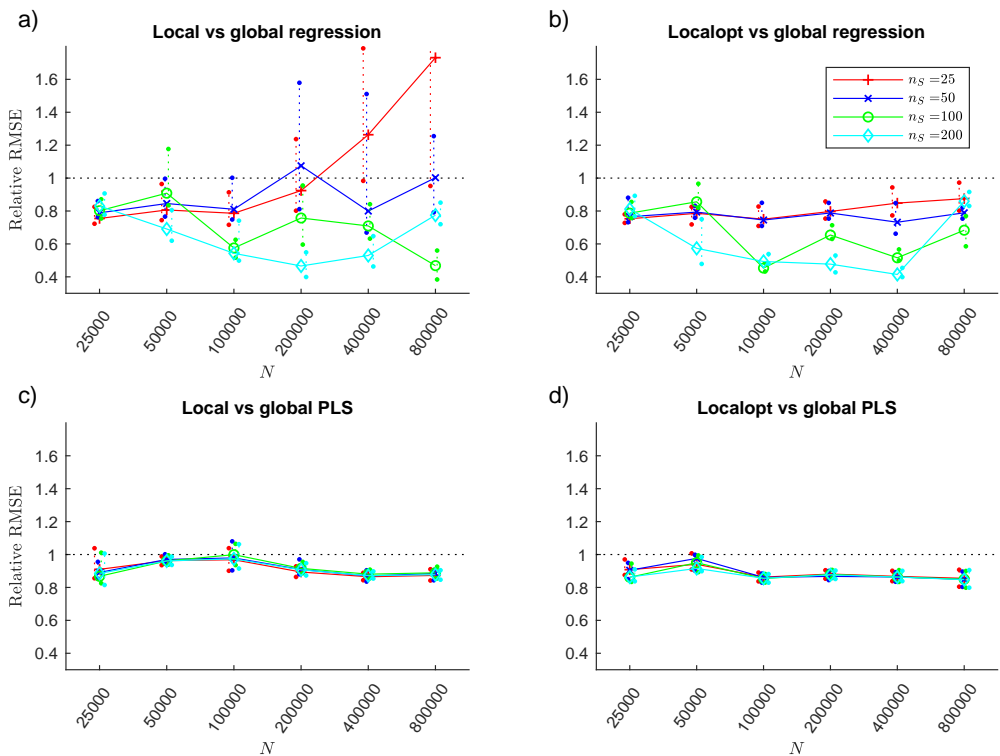}
	\caption{Reduction in RMSE using optimization and localization of the transformation with the $g$-and-$k$-distribution evaluated over simulated test datasets for different numbers of simulations and candidate summaries. The panels show the median relative SRMSEs over test datasets for local over global regression (a), optimized  local over global regression (b), local over global PLS (c) and optimized local over global PLS (d)  as a function of number of simulations ($N$). Each line shows the reduction for one number of candidate summaries ($n_S$) as indicated in the legend. The dotted vertical lines indicate 90 \% intervals for the SRMSEs over the test datasets.}
	\label{fig:gkRelRMSE}
\end{figure*}

Localized versions of PLS provided clear improvement in accuracy compared to their global counterparts with all combinations of numbers of simulations and candidate summaries (Fig. \ref{fig:gkRelRMSE}c,d). Optimization of the local PLS resulted in slightly higher accuracy over the regular version, but the difference was not large (Fig. S5c in the Supplementary material). With localized PLS the decrease in SRMSE was not affected by the number of candidate summaries, and was slightly negatively correlated with the number of simulations (Fig. \ref{fig:gkRelRMSE}c,d). 

Optimized local linear regression provided increased accuracy compared to global regression (Fig. \ref{fig:gkRelRMSE}b). With regular local linear regression there was in many cases improvement over global regression, but with a high number of simulations $N$ it sometimes resulted in poorer accuracy (Fig. \ref{fig:gkRelRMSE}a). With linear regression localization improved the accuracy most when the number of candidate summaries was high (100 or 200), with high variability between test datasets and combinations. The large decrease in SRMSE in high dimensions with the linear regression was caused by the failure of the global regression transformation (Fig. S6o-r,t-x in the Supplementary material). In the highest dimensional setting ($N=800,000$, $n_S=200$, Fig. S6s in the Supplementary material) even the localized versions of linear regression failed to produce useful summaries. This could be due to too high lower limit for the size of the local neighborhood $\alpha$, because $\alpha$ was optimized to the lower bound $0.0316$ for all test datasets (Fig. S7a in the Supplementary material). A smaller lower limit could have improved the results, but we did not investigate this. We chose the lower bound to produce reasonable localization even with the smallest number of simulations (25,000) and used the same candidates for all other numbers.

Overall, the size of the neighborhood $\alpha$ in the localized algorithms was smaller with regression than PLS, but dimensionality had only a small effect on it (Fig. S7 in the Supplementary material). Similarly, the number of components in different versions of PLS did not show a clear effect of dimensionality (Fig. S8 in the Supplementary material). The high variation between settings and datasets in the number of PLS components and quantiles for optimized local PLS indicates that there is usually not a single optimal value for these, but different combinations produced similar results.  The optimized PLS did not differ significantly from the regular PLS (Fig. S5b). 

The optimized local versions of both linear regression and PLS were clearly superior over the global versions for estimating $g$ and $k$, but provided similar or lower accuracy for $A$ and $B$ (Fig. S9  in the Supplementary material). This was probably caused by the optimized local algorithms minimizing sum of RMSEs over parameters. As there was higher uncertainty in $g$ and $k$, the poorer relative accuracy in $A$ and $B$ did not affect the overall accuracy as much.

The average running time for all of the dimension reduction methods scaled linearly with both the number of simulations and number of summaries (Fig. S10 in the Supplementary material). For most of the methods, the running time roughly doubled, when the either number of simulations or summaries doubled. Optimized local PLS was computationally significantly more expensive than the others due to the multidimensional optimization of both numbers of components and the size of the neighborhood. For regression the cost of optimizing the local transformation was lower than for PLS.

\section{Discussion}

We introduced a localization strategy for projection-based dimension reduction techniques for summary statistics. The introduced algorithm creates a low-dimensional transformation of the summaries, which is optimized to perform well in the neighborhood of the observed data. The proposed localization strategy is general and can be used with any projection-based dimension reduction technique to improve the efficiency of likelihood-free inference.

The optimization of the localized transformation over validation datasets guarantees good performance of the transformed summaries also outside the local neighborhood. This is in contrast to the similar localization strategy suggested earlier by \citet{AeschbacherBeaumontFutschik2012}, which did not validate or optimize the constructed summaries in the whole space. Our results show that the optimization improves the accuracy of summaries produced with localization. Although the difference is not large in many cases, the improvement is consistent and optimized localization provides at least as good accuracy as regular localization in almost all cases. More importantly, the optimization results in more stable transformations with less variation in accuracy among datasets, whereas localization without optimization sometimes produces poorly behaving summaries. The improvements provided by the optimization were more pronounced in high-dimensional situations, such as with the White-starred robin model, for which the non-optimized localization in many cases lead to inferior performance. The failure was probably caused by too small neighborhood used for localization. While this could be fixed by using a higher value for $\alpha$, it also highlights the need to adapt the dimension reduction technique to the problem at hand, which is automatically provided by our approach. For example, optimal value for $\alpha$ could depend on the number of simulations available, dimensionality of the summaries and the parameters, model structure and the observed data, making it difficult to know beforehand how large $\alpha$ should be.

Our results show that optimized localization is generally a preferred strategy over no localization or localization without optimization. However, the optimization as presented in Algorithm \ref{al:optlocaltransformation} does add a possibly significant computational cost to the inference. As the optimization is based on an exhaustive search over candidate transformation parameters $\varLambda$, a total of $N_{valid} |\varLambda|$  transformations have to be constructed instead of a single global transformation. In the case of transformation $f$ depending on parameters, such as the number of components in PLS, there would be multiple parameters to optimize and hence the size of $\varLambda$ would have to be high. Therefore, localization of a parameterized transformation might not be sensible for models from which it is very fast to simulate new datasets, such as the $g$-and-$k$-distribution and Ricker map. For computationally heavy models, for which simulation of one dataset could take minutes or even hours, localization helps to achieve higher accuracy without too big additional computational cost. However, the computational cost could easily be reduced by using more sophisticated optimization algorithms such as Bayesian optimization \citep{SnoekLarochelleAdams2012}, which could find the optimal solution with fewer parameter value $\lambda$ evaluations. Additionally, as the computational cost of the dimension reduction methods scales directly with the number of samples, the improvements provided by increasing the sample size and more complex dimension reduction methods may be directly compared.

The choice of the initial transformation $f_1$ used for localization in Algorithm \ref{al:optlocaltransformation} may have a significant impact on the performance of the transformed summaries. In most cases the use of the global version of the projection-based method works well, as shown by our results, but sometimes the global transformation could produce summaries that lead the localization to a wrong direction. For example, in the White-starred robin model global regression resulted in summaries under which zero simulations were closer to the test datasets than most of the positive simulations. As a result, the local transformations were mostly based on zero simulations and failed to be informative about the parameters. By using all the candidate summaries as initial summary statistics, the local regression worked as expected and provided a clear improvement over the global regression. While this kind of pathological performance is not expected to be common, it is still advisable to check that the initial transformation and the localized summaries are producing reasonable results. Unfortunately there does not exist any direct way of ensuring that the localization is working as expected, but insights may be obtained by comparing the global and localized summaries with simulated test datasets. On the other hand, the significant increase in accuracy with the White-starred robin model after localization, using all candidates, suggests that the initial choice does not need to be perfect. ABC is generally not expected to work well due to curse of dimensionality with over 300 summaries, but for localization of the final transformation their efficiency was adequate.

The example cases analysed in this work show interesting results concerning localized linear regression and PLS as dimension reduction techniques. When dimensionality was high, linear regression sometimes failed to produce good transformed summaries. The failure of regression occured with multiple models, whereas PLS seemed to work more robustly regardless of the problem. With the $g$-and-$k$ distribution, increase of both the number of simulations $N$ and number of summaries $n_S$ caused difficulties for regression. With the White-starred robin model, and to a lesser degree with the Ricker map, the cause of problems for regression was the high number of simulations with zero observations. The latter failure could be related to the differing principles behind the two methods. The zero simulation did not have an influence on the PLS components that were used as the transformed summaries with PLS. On the other hand, with linear regression the transformed summaries were the linear projections, which were distorted by the high-number of zero simulations. Having said that, we acknowledge that comparison between the two methods was not the main goal of the present work, and the results should be considered at most as suggestive of their general performance. Our results are also somewhat in contrast to the findings of \citet{BlumNunesPrangleEtAl2012}, who found that linear regression generally outperformed PLS and the difference was highest in high-dimensional settings. The relative performance of the methods seems to vary with the model under study, and more research on the subject would be needed to understand it better.

In this work we have focused on finding relatively simple linear transformations for the summaries. Such transformations are perhaps ideally suited for localization, as the linearity assumption is not expected to hold globally in the space of possible summaries, but locally within a restricted range linearity often is a reasonable approximation. Additionally, linear mappings are easy to learn even in high-dimensional settings allowing more narrow localization. However, the localization algorithms presented here are suitable for any other projection-based dimension reduction method, including non-linear regression approaches such as feed-forward neural networks \citep{BlumFrancois2010} or boosting \citep{AeschbacherBeaumontFutschik2012}. With the more complicated transformations localization might not lead to as big improvements, since they require a larger number of samples to be fit and, at the same time, might capture the true relationship better in a larger neighborhood. Whether a more narrow linear transformation provides in general better summaries than a wider non-linear transformation remains an open question, although the comparisons in \citet{BlumNunesPrangleEtAl2012} suggest that non-linear methods lead to roughly similar performance as the linear methods. The transformation of the summaries produces a scaling for the candidate summaries, and for the accuracy of the ABC inference it mostly matters in the neighborhood around the analyzed dataset. If more complex transformations provide better scaling further away from the data, it might not have any significant impact on the inference results.

All dimension reduction techniques for summary statistics, including the one introduced in this work, are based on rejection sampling ABC, which is computationally inefficient in anything but low-dimensional problems. The main advantage of rejection sampling is that it facilitates performing multiple ABC analyses for optimizing the dimension reduction using the same set of simulations, which would not be possible with any other ABC algorithm. If the accuracy provided by the rejection sampling is not enough, the transformed summaries may be used in another ABC analysis with a more advanced ABC algorithm, such as ABC SMC \citep{ToniWelchStrelkowaEtAl2009} or BOLFI \citep{GutmannCorander2015}. However, it might be possible to extend SMC-type ABC algorithms to simultaneously target the posterior and adapt the transformation of the summaries. \citet{Prangle2017} introduced a population Monte Carlo ABC algorithm that adapted the weights of the summaries in the distance function at every step, and mentioned the possibility that it could be extended to adapt the summaries themselves. Ensuring convergence of such an algorithm might prove to be difficult, because the algorithm would at the same time be modifying the target and trying to concentrate locally around the target. Nevertheless, such an algorithm could provide a significant increase in efficiency for ABC analysis of models with a high number of candidate summaries, and hence more research on the subject would be justified.

\begin{acknowledgements}
We acknowledge the computational resources provided by the Aalto Science-IT project and support by the Finnish Center for Artificial Intelligence Research FCAI. The work was supported by the Academy of Finland (grants 319264, 294238 and 292334).
\end{acknowledgements}

\bibliographystyle{spbasic}      
\bibliography{refs}   

\begin{thebibliography}{26}
\providecommand{\natexlab}[1]{#1}
\providecommand{\url}[1]{{#1}}
\providecommand{\urlprefix}{URL }
\expandafter\ifx\csname urlstyle\endcsname\relax
  \providecommand{\doi}[1]{DOI~\discretionary{}{}{}#1}\else
  \providecommand{\doi}{DOI~\discretionary{}{}{}\begingroup
  \urlstyle{rm}\Url}\fi
\providecommand{\eprint}[2][]{\url{#2}}

\bibitem[{Aeschbacher et~al.(2012)Aeschbacher, Beaumont, and
  Futschik}]{AeschbacherBeaumontFutschik2012}
Aeschbacher S, Beaumont MA, Futschik A (2012) A novel approach for choosing
  summary statistics in approximate {Bayesian} computation. Genetics
  192(3):1027--1047

\bibitem[{Barber et~al.(2015)Barber, Voss, and Webster}]{BarberVossWebster2015}
Barber S, Voss J, Webster M (2015) The rate of convergence for approximate
  bayesian computation. Electron J Statist 9(1):80--105,
  \doi{10.1214/15-EJS988}, \urlprefix\url{http://dx.doi.org/10.1214/15-EJS988}

\bibitem[{Beaumont et~al.(2002)Beaumont, Zhang, and
  Balding}]{BeaumontZhangBalding2002}
Beaumont MA, Zhang W, Balding DJ (2002) Approximate {Bayesian} computation in
  population genetics. Genetics 162(4):2025--2035

\bibitem[{Blum et~al.(2013)Blum, Nunes, Prangle, and
  Sisson}]{BlumNunesPrangleEtAl2012}
Blum M, Nunes M, Prangle D, Sisson S (2013) A comparative review of dimension
  reduction methods in approximate {Bayesian} computation. Statistical Science
  28(2):189--208

\bibitem[{Blum and Fran{\c{c}}ois(2010)}]{BlumFrancois2010}
Blum MGB, Fran{\c{c}}ois O (2010) Non-linear regression models for approximate
  {Bayesian} computation. Statistics and Computing 20(1):63--73

\bibitem[{Drovandi et~al.(2015)Drovandi, Pettitt, and
  Lee}]{DrovandiPettittLee2015}
Drovandi CC, Pettitt AN, Lee A (2015) Bayesian indirect inference using a
  parametric auxiliary model. Statistical Science 30:72--95

\bibitem[{Fearnhead and Prangle(2012)}]{FearnheadPrangle2012}
Fearnhead P, Prangle D (2012) Constructing summary statistics for approximate
  {Bayesian} computation: semi-automatic approximate {Bayesian} computation.
  Journal of the Royal Statistical Society Series B (Methodological) 74:1--28

\bibitem[{Gutmann and Corander(2016)}]{GutmannCorander2015}
Gutmann MU, Corander J (2016) Bayesian optimization for likelihood-free
  inference of simulator-based statistical models. The Journal of Machine
  Learning Research 17(1):4256--4302

\bibitem[{Gutmann et~al.(2018)Gutmann, Dutta, Kaski, and
  Corander}]{GutmannDuttaKaskiEtAl2014}
Gutmann MU, Dutta R, Kaski S, Corander J (2018) Likelihood-free inference via
  classification. Statistics and Computing 28(2):411--425

\bibitem[{Haynes et~al.(1997)Haynes, MacGillivray, and
  Mengersen}]{HaynesMacGillivrayMengersen1997}
Haynes MA, MacGillivray H, Mengersen K (1997) Robustness of ranking and
  selection rules using generalised g-and-k distributions. Journal of
  Statistical Planning and Inference 65(1):45--66

\bibitem[{Joyce and Marjoram(2008)}]{JoyceMarjoram2008}
Joyce P, Marjoram P (2008) Approximately sufficient statistics and {Bayesian}
  computation. Statistical Applications in Genetics and Molecular Biology 7(1)

\bibitem[{Kangasr\"{a}\"{a}si\"{o} et~al.(2017)Kangasr\"{a}\"{a}si\"{o},
  Athukorala, Howes, Corander, Kaski, and
  Oulasvirta}]{KangasraeaesioeAthukoralaHowesEtAl2017}
Kangasr\"{a}\"{a}si\"{o} A, Athukorala K, Howes A, Corander J, Kaski S,
  Oulasvirta A (2017) Inferring cognitive models from data using approximate
  {Bayesian} computation. In: Proceedings of the 2017 CHI Conference on Human
  Factors in Computing Systems, CHI '17, ACM, New York, NY, USA, pp 1295--1306,
  \doi{10.1145/3025453.3025576},
  \urlprefix\url{http://doi.acm.org/10.1145/3025453.3025576}

\bibitem[{Lintusaari et~al.(2017)Lintusaari, Gutmann, Dutta, Kaski, and
  Corander}]{LintusaariGutmannDuttaEtAl2016}
Lintusaari J, Gutmann MU, Dutta R, Kaski S, Corander J (2017) Fundamentals and
  recent developments in approximate {Bayesian} computation. Systematic Biology
  66(1):e66--e82

\bibitem[{Lintusaari et~al.(2018)Lintusaari, Vuollekoski,
  Kangasr{\"a}{\"a}si{\"o}, Skyt{\'e}n, J{\"a}rvenp{\"a}{\"a}, Marttinen,
  Gutmann, Vehtari, Corander, and
  Kaski}]{LintusaariVuollekoskiKangasraeaesioeEtAl2017}
Lintusaari J, Vuollekoski H, Kangasr{\"a}{\"a}si{\"o} A, Skyt{\'e}n K,
  J{\"a}rvenp{\"a}{\"a} M, Marttinen P, Gutmann MU, Vehtari A, Corander J,
  Kaski S (2018) {ELFI}: engine for likelihood-free inference. The Journal of
  Machine Learning Research 19(1):643--649

\bibitem[{Nunes and Balding(2010)}]{NunesBalding2010}
Nunes M, Balding D (2010) On optimal selection of summary statistics for
  approximate {Bayesian} computation. Statistical Applications in Genetics and
  Molecular Biology 9(1):34

\bibitem[{Peters et~al.(2018)Peters, Panayi, and
  Septier}]{PetersPanayiSeptier2018}
Peters GW, Panayi E, Septier F (2018) Sequential {Monte Carlo}-{ABC} methods
  for estimation of stochastic simulation models of the limit order book. In:
  Handbook of Approximate Bayesian Computation, Chapman and Hall/CRC, pp
  437--480

\bibitem[{Prangle(2017)}]{Prangle2017}
Prangle D (2017) Adapting the {ABC} distance function. Bayesian Anal
  12(1):289--309

\bibitem[{Prangle(2018)}]{Prangle2015a}
Prangle D (2018) Summary statistics in approximate {Bayesian} computation. In:
  Sisson S, Fan Y, Beaumont M (eds) Handbook of Approximate Bayesian
  Computation, Chapman and Hall/CRC

\bibitem[{Pritchard et~al.(1999)Pritchard, Seielstad, Perez-Lezaun, and
  Feldman}]{PritchardSeielstadPerez-LezaunEtAl1999}
Pritchard JK, Seielstad MT, Perez-Lezaun A, Feldman MW (1999) Population growth
  of human {Y} chromosomes: a study of {Y} chromosome microsatellites.
  Molecular Biology and Evolution 16(12):1791--1798

\bibitem[{Sirén et~al.(2018)Sirén, Lens, Cousseau, and
  Ovaskainen}]{SirenLensCousseauEtAl2018}
Sirén J, Lens L, Cousseau L, Ovaskainen O (2018) Assessing the dynamics of
  natural populations by fitting individual based models with approximate
  {Bayesian} computation. Methods in Ecology and Evolution 9(5):1286--1295

\bibitem[{{Sisson} et~al.(2018){Sisson}, {Fan}, and
  {Beaumont}}]{SissonFanBeaumont2018}
{Sisson} SA, {Fan} Y, {Beaumont} MA (2018) {Overview of Approximate Bayesian
  Computation}. In: Handbook of Approximate Bayesian Computation, Chapman and
  Hall/CRC

\bibitem[{Snoek et~al.(2012)Snoek, Larochelle, and
  Adams}]{SnoekLarochelleAdams2012}
Snoek J, Larochelle H, Adams RP (2012) Practical {Bayesian} optimization of
  machine learning algorithms. In: Pereira F, Burges C, Bottou L, Weinberger K
  (eds) Advances in Neural Information Processing Systems 25, Curran
  Associates, Inc., pp 2951--2959

\bibitem[{Tavaré et~al.(1997)Tavaré, Balding, Griffiths, and
  Donnelly}]{TavareBaldingGriffithsEtAl1997}
Tavaré S, Balding DJ, Griffiths RC, Donnelly P (1997) Inferring coalescence
  times from {DNA} sequence data. Genetics 145(2):505--518

\bibitem[{Toni et~al.(2009)Toni, Welch, Strelkowa, Ipsen, and
  Stumpf}]{ToniWelchStrelkowaEtAl2009}
Toni T, Welch D, Strelkowa N, Ipsen A, Stumpf MP (2009) Approximate {Bayesian}
  computation scheme for parameter inference and model selection in dynamical
  systems. Journal of the Royal Society Interface 6(31):187--202

\bibitem[{Wegmann et~al.(2009)Wegmann, Leuenberger, and
  Excoffier}]{WegmannLeuenbergerExcoffier2009}
Wegmann D, Leuenberger C, Excoffier L (2009) Efficient approximate {Bayesian}
  computation coupled with {Markov} chain {Monte} {Carlo} without likelihood.
  Genetics 182(4):1207--1218

\bibitem[{Wood(2010)}]{Wood2010}
Wood SN (2010) Statistical inference for noisy nonlinear ecological dynamic
  systems. Nature 466(7310):1102--1104

\end{thebibliography}

\end{document}


\raggedright

	{\Large
		\textbf{Supplementary material for Local dimension reduction of summary statistics for likelihood-free inference}

	}
	Jukka Sir\'{e}n, 
	Samuel Kaski

\section*{Supplementary tables}

\begin{center}
\label{tab:avalfa}
\captionof{table}{The average size of local neighborhood $\alpha$ used for learning localized transformation with different models over test datasets. 90 \% intervals for $\alpha$ over the test datasets are shown in brackets. 'localRegopt' and 'localPLSopt' refer to optimized local versions of regression and PLS, respectively.}
\begin{tabular}{lll}
	\hline\noalign{\smallskip} 
	& localRegopt & localPLSopt  \\ 
	\hline\noalign{\smallskip}
	Ricker &  0.039 (0.032,0.076) & 0.104 (0.032,0.355) \\ 
	WR IBM: all  & 0.112 (0.045,0.355) & 0.159 (0.067,0.214)  \\ 
	WR IBM: pos  & 0.088 (0.045,0.126) & 0.041 (0.032,0.089) \\ 
	\hline\noalign{\smallskip}
\end{tabular} 
\end{center}

\begin{center}
	\label{tab:avPLS}
	\captionof{table}{The average number of PLS components used with different models and under different dimension reduction strategies over 100 test datasets. 90 \% intervals for the number of components over the test datasets are shown in brackets, when applicable. 'PLS' and  'PLSopt' refer to regular and optimized PLS transformations. 'localPLS' and 'localPLSopt' refer to local versions of PLS transformation without (Algorithm 1) and with optimization (Algorithm 2), respectively. 'initial' refers to the first transformation and 'local' to the second transformation in the local transformations. For 'localPLS' the number of components in the first transformation is the same as for 'PLS'.}
	\begin{tabular}{llllll} 
		\hline\noalign{\smallskip}
		& PLS & PLSopt & localPLS: local & localPLSopt: initial & localPLSopt: local \\ 
		\hline\noalign{\smallskip}
		Ricker &  8 & 9.2 (4,11) & 7.3 (6,9) & 5.9 (2,14.5) & 4.6 (3,8) \\ 
		WR IBM: all & 7 & 14 (14,14)  & 15 (15,15) & 1.56 (1,4) & 13.5 (11,15) \\
		WR IBM: pos & 13 & 13.6 (12,15) & 15 (15,15) & 12.7 (11,15) & 13.4 (12,15)\\
		\hline\noalign{\smallskip}
	\end{tabular} 
\end{center}

\newpage

\section*{Supplementary figures}

\begin{center}
	\includegraphics[width=18cm]{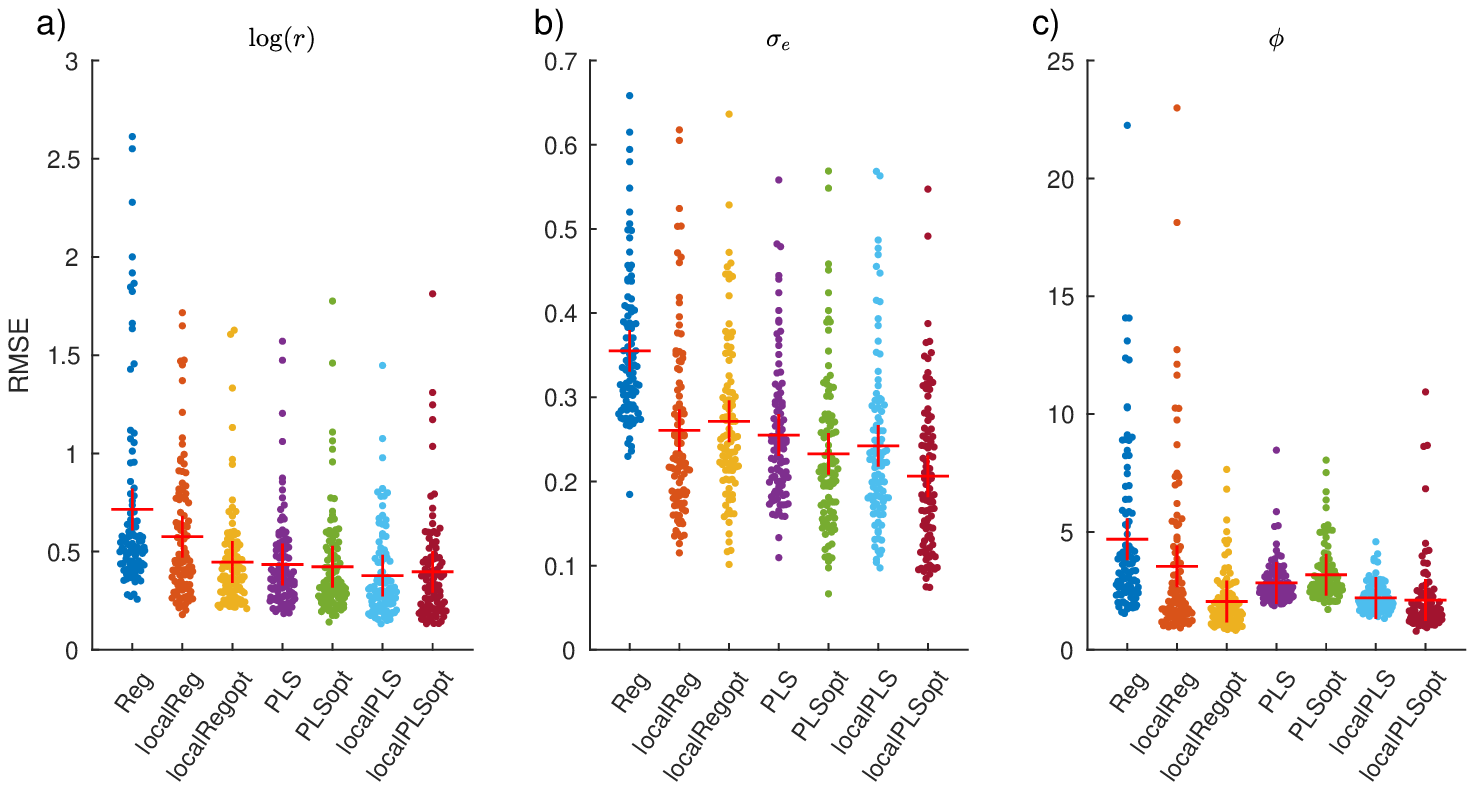}
	\captionof{figure}{Accuracy of different dimension reduction techniques with the Ricker model evaluated over simulated test datasets. Each panel shows the RMSEs for one parameter with each distribution showing the average RMSE for the 100 test datesets. 'Reg' and 'PLS' refer to global regression and PLS transformations for the parameters. 'PLSopt' refers to PLS with the number of components optimized using validation datasets. 'localReg' and 'localPLS' refer to local versions of the regression and PLS transformation with Algorithm 1, respectively. 'localRegopt' and 'localPLSopt' refer to optimized local versions of the regression and PLS transformation with Algorithm 2, respectively.}
	\label{fig:rickerRMSEall}
\end{center}

\begin{center}
	\includegraphics[width=18cm]{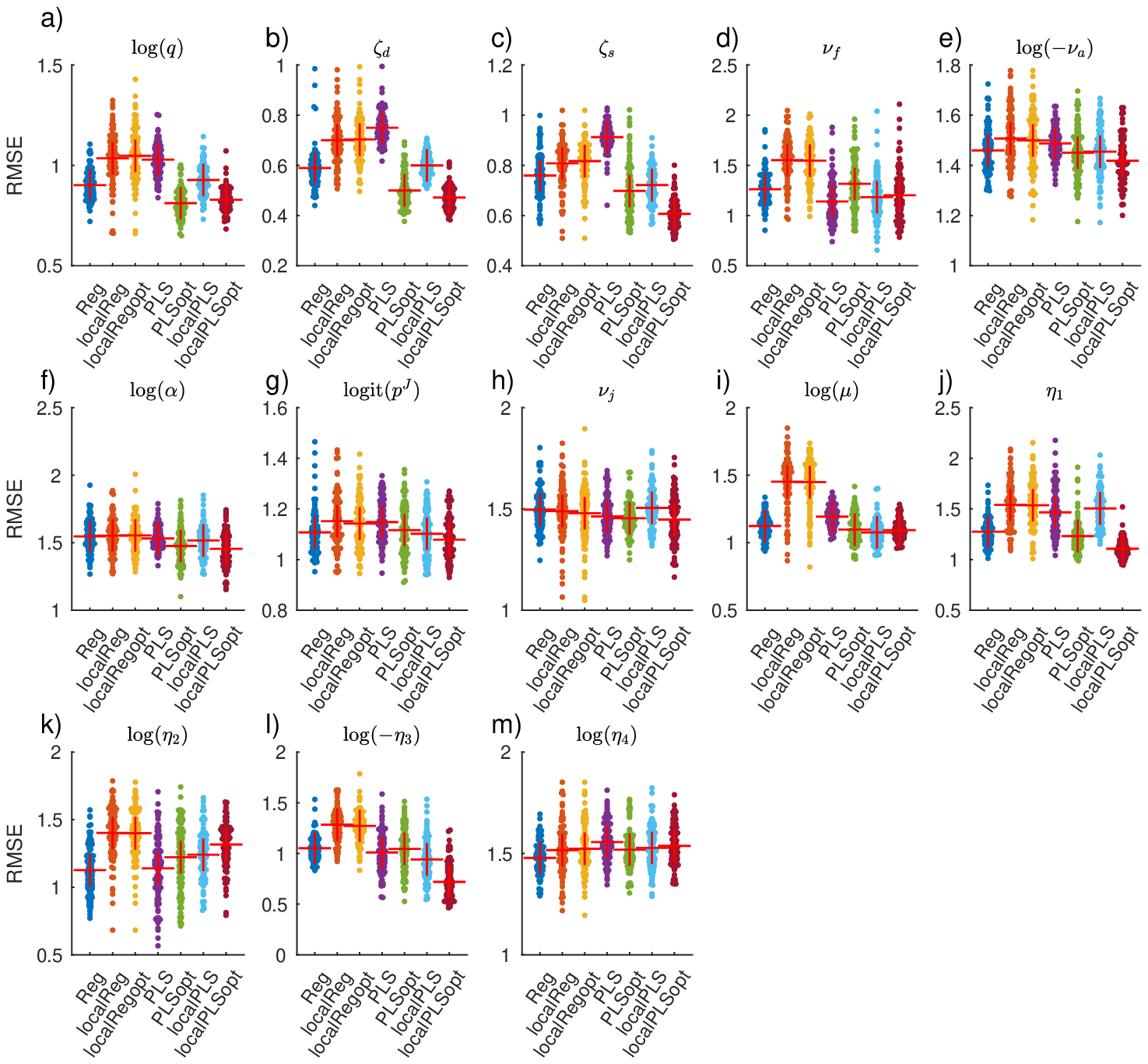}
	\captionof{figure}{Accuracy of different dimension reduction techniques with the bird population dynamics IBM evaluated over simulated test datasets using all simulations. Each panel shows the RMSEs for one parameter with each distribution showing the average RMSE for the 100 test datesets. 'Reg' and 'PLS' refer to global regression and PLS transformations for the parameters. 'PLSopt' refers to PLS with the number of components optimized using validation datasets. 'localReg' and 'localPLS' refer to local versions of the regression and PLS transformation with Algorithm 1, respectively. 'localRegopt' and 'localPLSopt' refer to optimized local versions of the regression and PLS transformation with Algorithm 2, respectively.}
	\label{fig:WRRMSEall}
\end{center}

\begin{center}
	\includegraphics[width=18cm]{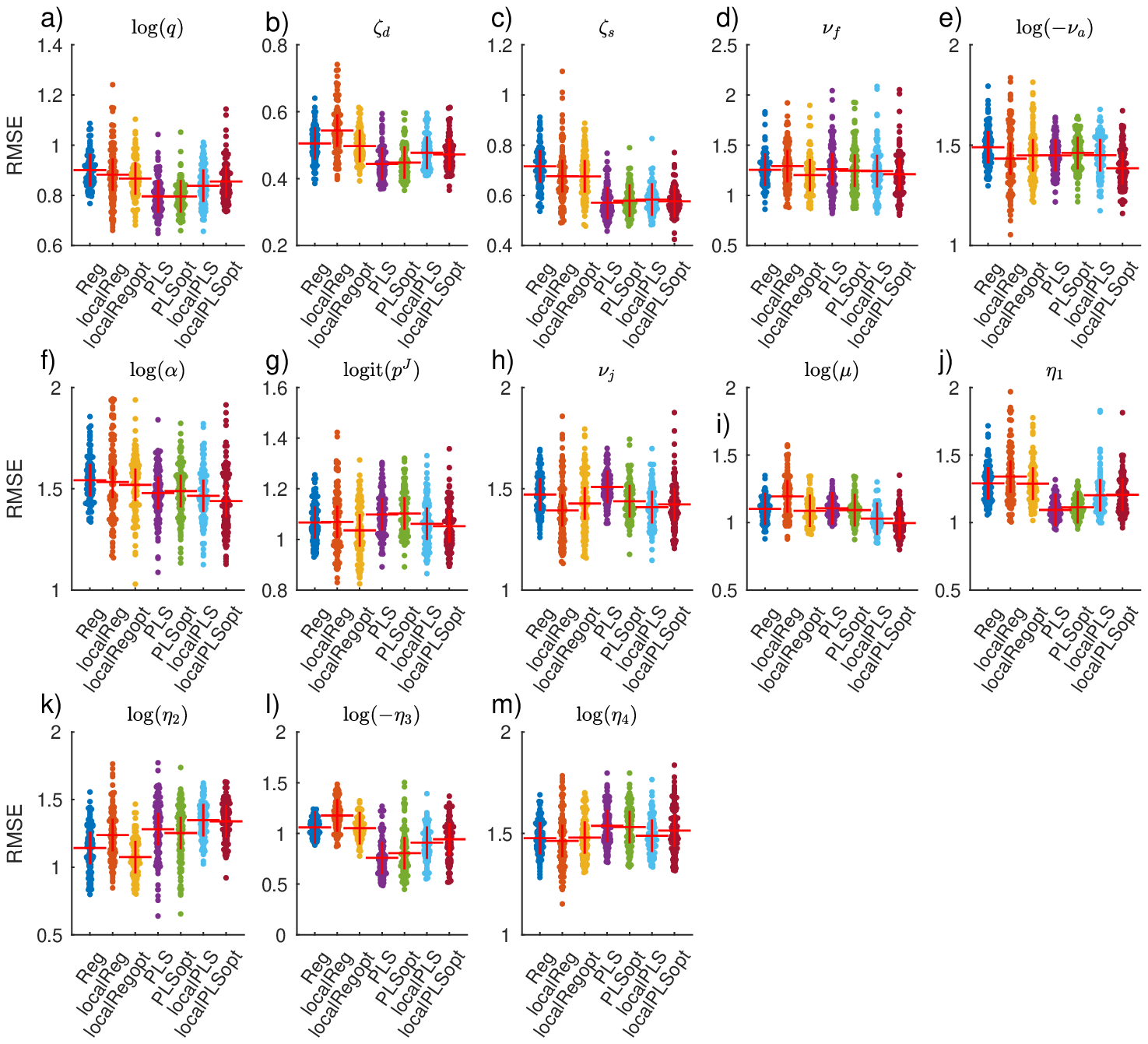}
	\captionof{figure}{Accuracy of different dimension reduction techniques with the bird population dynamics IBM evaluated over simulated test datasets using only positive simulations. Each panel shows the RMSEs for one parameter with each distribution showing the average RMSE for the 100 test datesets. 'Reg' and 'PLS' refer to global regression and PLS transformations for the parameters. 'PLSopt' refers to PLS with the number of components optimized using validation datasets. 'localReg' and 'localPLS' refer to local versions of the regression and PLS transformation with Algorithm 1, respectively. 'localRegopt' and 'localPLSopt' refer to optimized local versions of the regression and PLS transformation with Algorithm 2, respectively.}
	\label{fig:WRn0RMSEall}
\end{center}

\begin{center}
	\includegraphics[width=18cm]{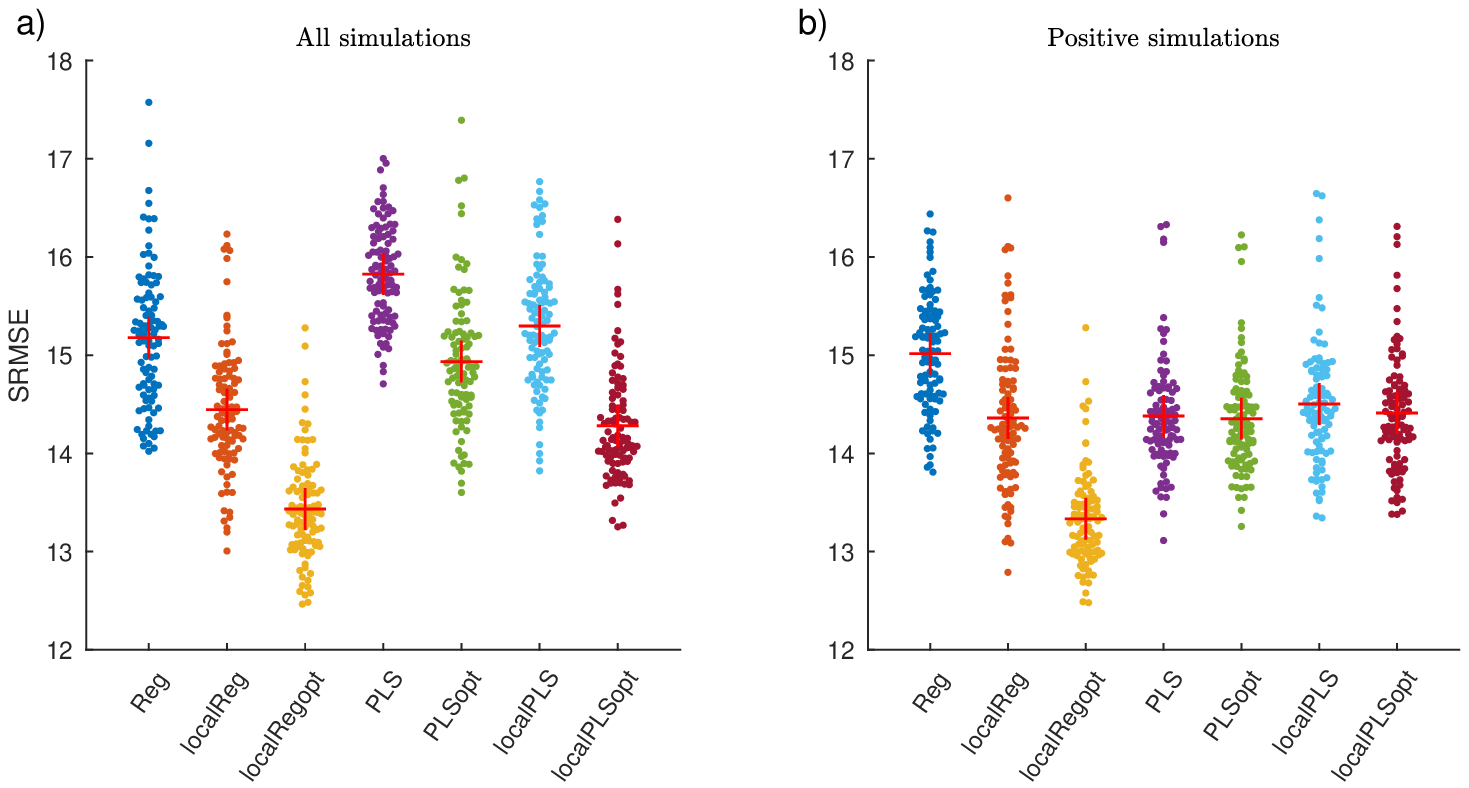}
	\captionof{figure}{Same as Fig 2, but with identity function instead of global regression as the initial transformation for the local regressions ('localReg' and 'localRegopt'). Accuracy of different dimension reduction techniques with the bird population dynamics IBM evaluated over simulated test datasets. The plots shows the average SRMSE for the 100 test datesets using all simulations (a or using only those 25,170 simulations with observations (b). 'Reg' and 'PLS' refer to global regression and PLS transformations for the parameters. 'PLSopt' refers to PLS with the number of components optimized using validation datasets.'localReg' and 'localPLS' refer to local versions of the regression and PLS transformation with Algorithm 1, respectively. 'localRegopt' and 'localPLSopt' refer to optimized local versions of the regression and PLS transformation with Algorithm 2, respectively.}
	\label{fig:WRRMSEsumb}
\end{center}

\newpage

\begin{center}
	\includegraphics[width=18cm]{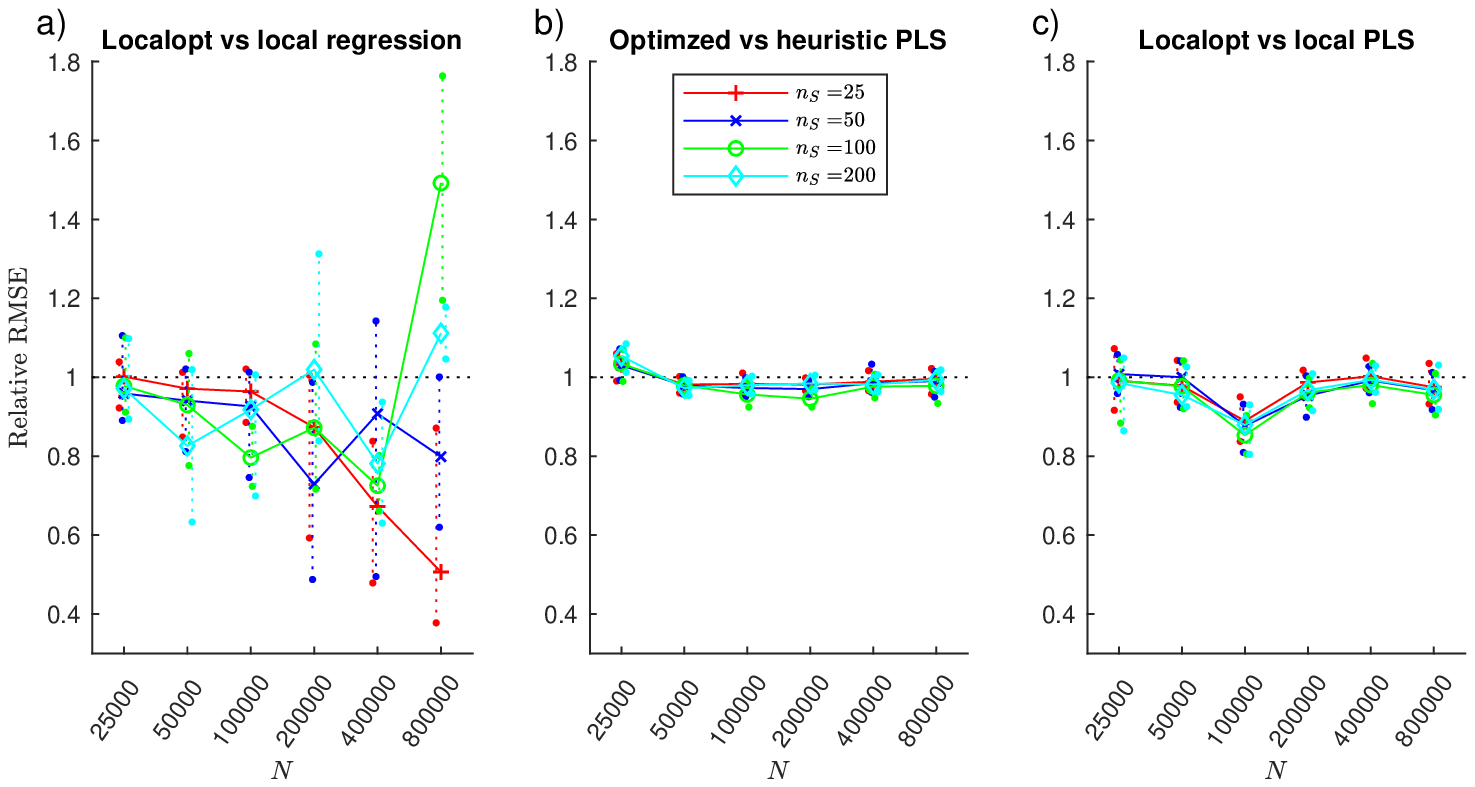}
	\captionof{figure}{Reduction in RMSE using optimization and localization of the transformation with the $g$-and-$k$-distribution evaluated over simulated test datasets for different numbers of simulations and candidate summaries. The panels show the median relative SRMSEs over test datasets for optimized over regular local regression (a), optimized over regular PLS (b) and optimized over regular local PLS (c)  as a function of number of simulations ($N$). Each line shows the reduction for one number of candidate summaries ($n_S$) as indicated in the legend. The dotted vertical lines indicate 90 \% intervals for the SRMSEs over the test datasets.}
	\label{fig:gkRelRMSE_supp}
\end{center}

\newpage

\begin{center}
	\includegraphics[width=18cm]{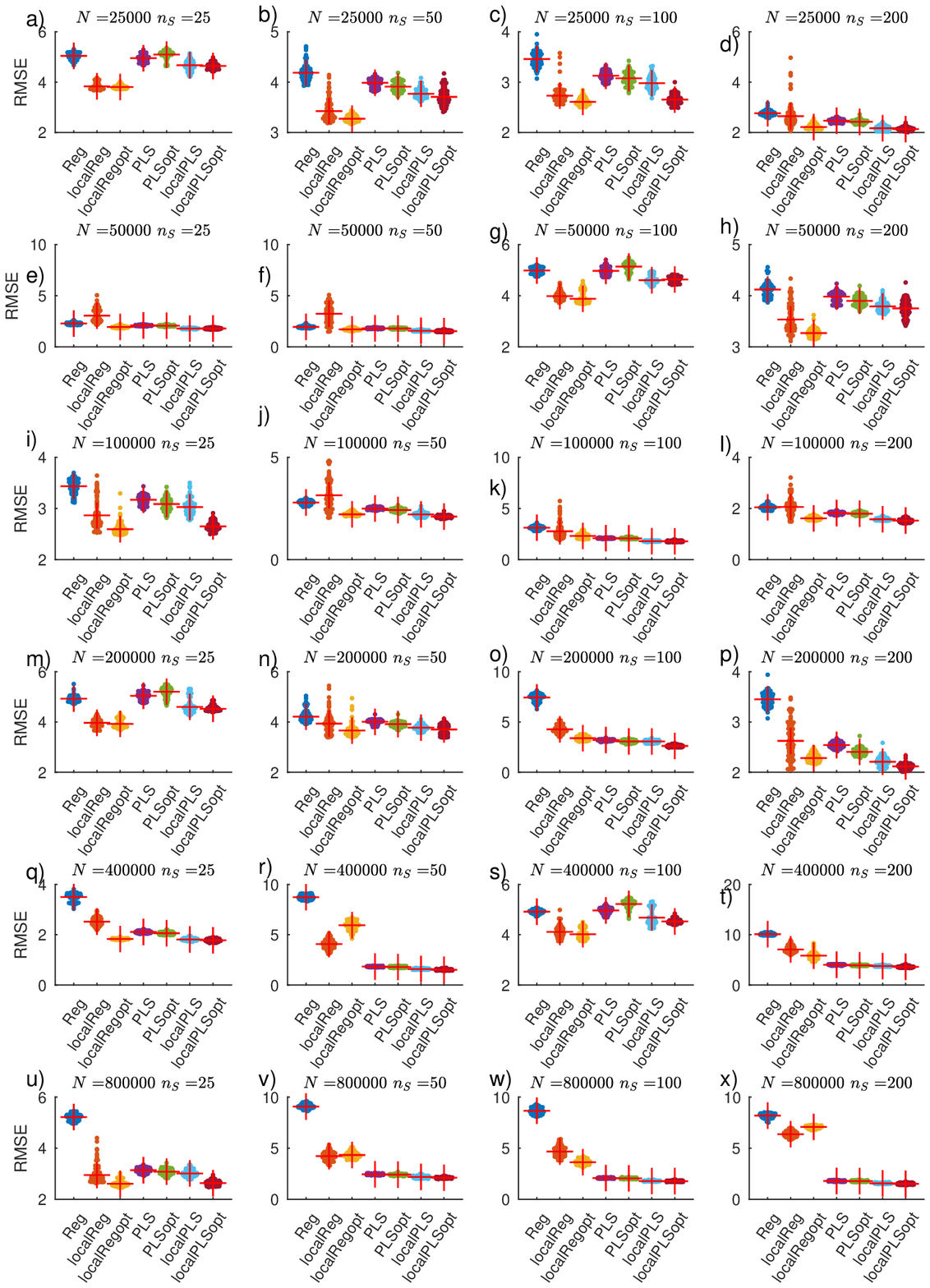}
	\captionof{figure}{Accuracy of different dimension reduction techniques with the $g$-and-$k$-distribution evaluated over simulated test datasets for different numbers of simulations and candidate summaries. Each panel shows the sum of RMSEs over the parameter with each distribution showing the average RMSE for the 100 test datesets for one combination of number of simulations ($N$) and number of candidate summaries ($n_S$). 'Reg' and 'PLS' refer to global regression and PLS transformations for the parameters. 'PLSopt' refers to PLS with the number of components optimized using validation datasets. 'localReg' and 'localPLS' refer to local versions of the regression and PLS transformation with Algorithm 1, respectively. 'localRegopt' and 'localPLSopt' refer to optimized local versions of the regression and PLS transformation with Algorithm 2, respectively.}
	\label{fig:gkRMSEdim}
\end{center}

\begin{center}
	\includegraphics[width=18cm]{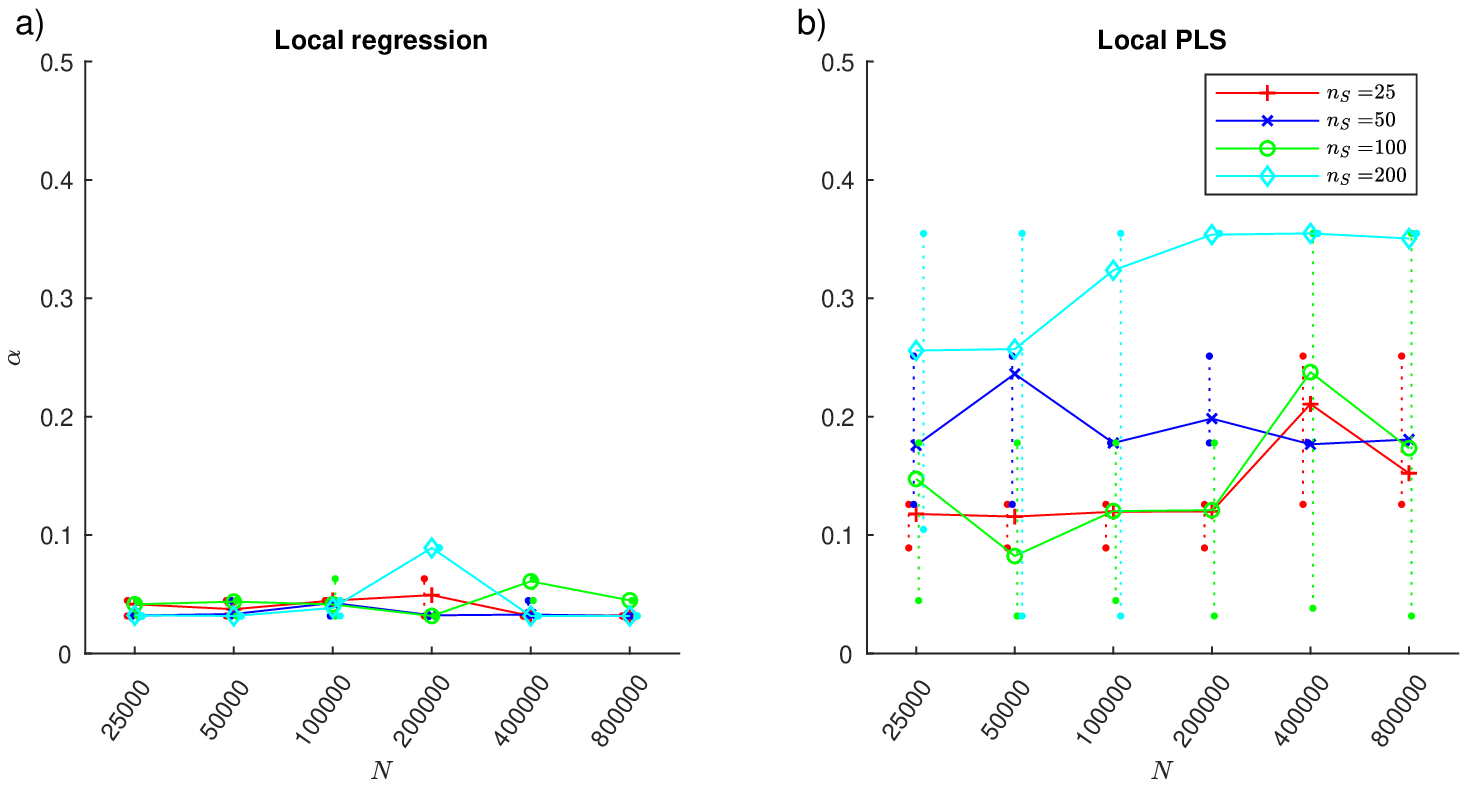}
	\captionof{figure}{The size of local neighborhood $\alpha$ used for learning localized transformation with the $g$-and-$k$-distribution evaluated over simulated test datasets for different numbers of simulations and candidate summaries. The panels show the average $\alpha$ over test datasets for optimized local regression (a) and optimized local PLS (b). The dotted vertical lines indicate 90 \% intervals for $\alpha$ over the test datasets.}
	\label{fig:gkalpha}
\end{center}

\newpage

\begin{center}
	\includegraphics[width=18cm]{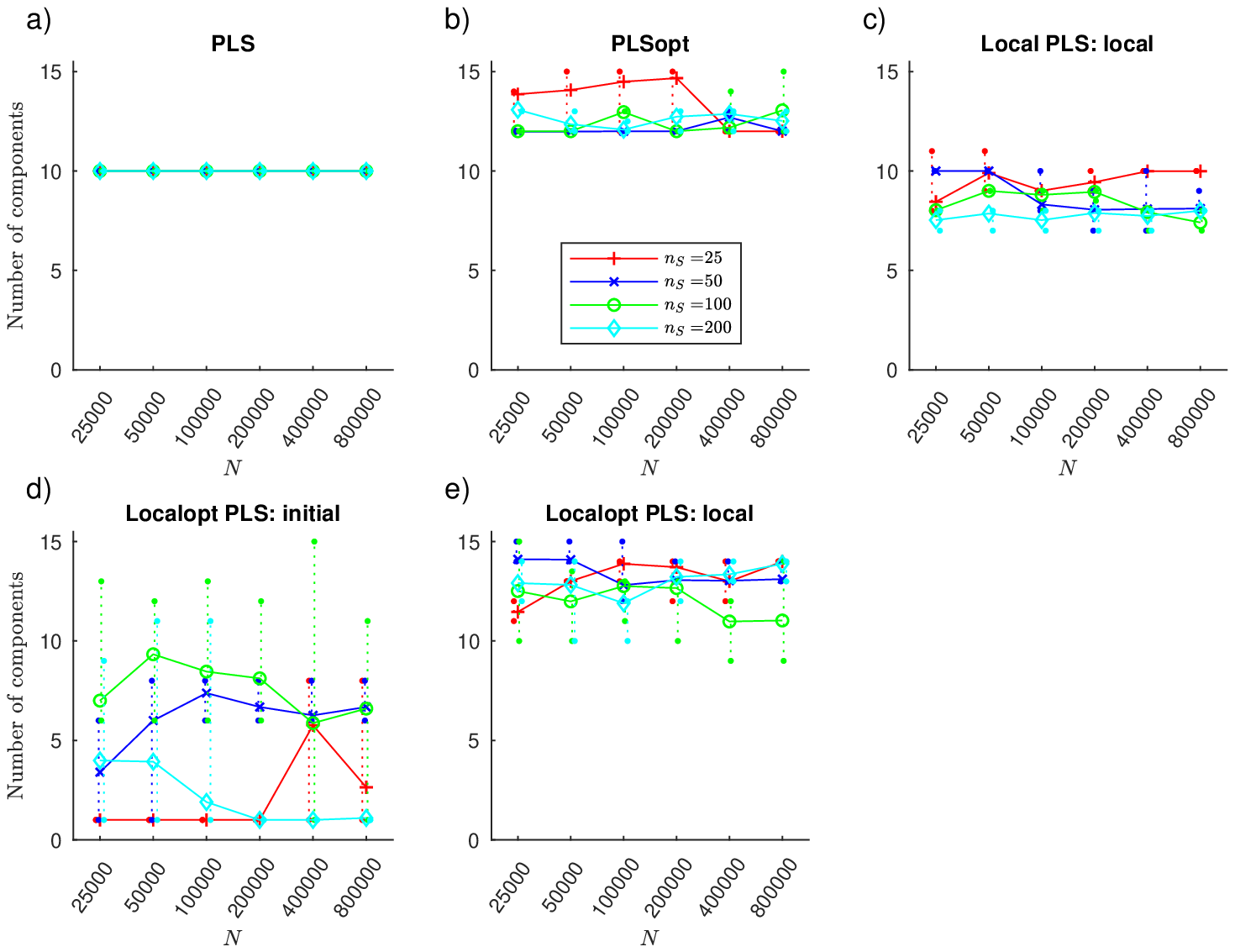}
	\captionof{figure}{The number of components used for PLS with the $g$-and-$k$-distribution evaluated over simulated test datasets for different numbers of simulations and candidate summaries. The panels show the average number of components over test datasets for regular PLS (a), optimized PLS (b), local transformation in regular local PLS (d), initial transformation in optimized local PLS (d) and local transformation in optimized local PLS (e). The dotted vertical lines indicate 90 \% intervals for the number of components over the test datasets.}
	\label{fig:gkncomp}
\end{center}

\newpage

\begin{center}
	\includegraphics[width=18cm]{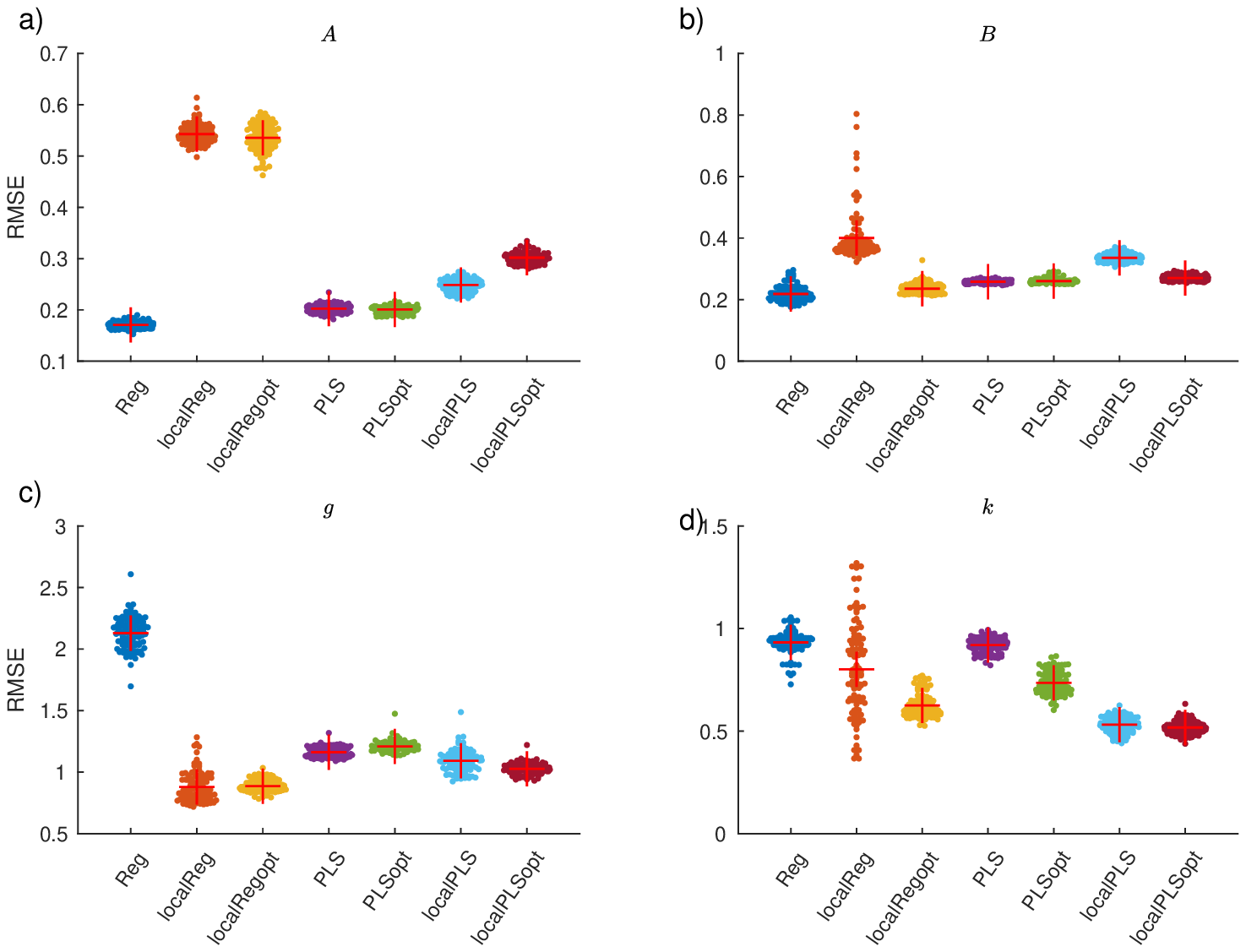}
	\captionof{figure}{Accuracy of different dimension reduction techniques with the $g$-and-$k$-distribution evaluated over simulated test datasets for the setting with 100 candidate summaries and 200,000 simulations. Each panel shows the RMSEs for one parameter with each distribution showing the average RMSE for the 100 test datesets. 'Reg' and 'PLS' refer to global regression and PLS transformations for the parameters. 'PLSopt' refers to PLS with the number of components optimized using validation datasets. 'localReg' and 'localPLS' refer to local versions of the regression and PLS transformation with Algorithm 1, respectively. 'localRegopt' and 'localPLSopt' refer to optimized local versions of the regression and PLS transformation with Algorithm 2, respectively.}
	\label{fig:gkRMSEall}
\end{center}

\begin{center}
	\includegraphics[width=18cm]{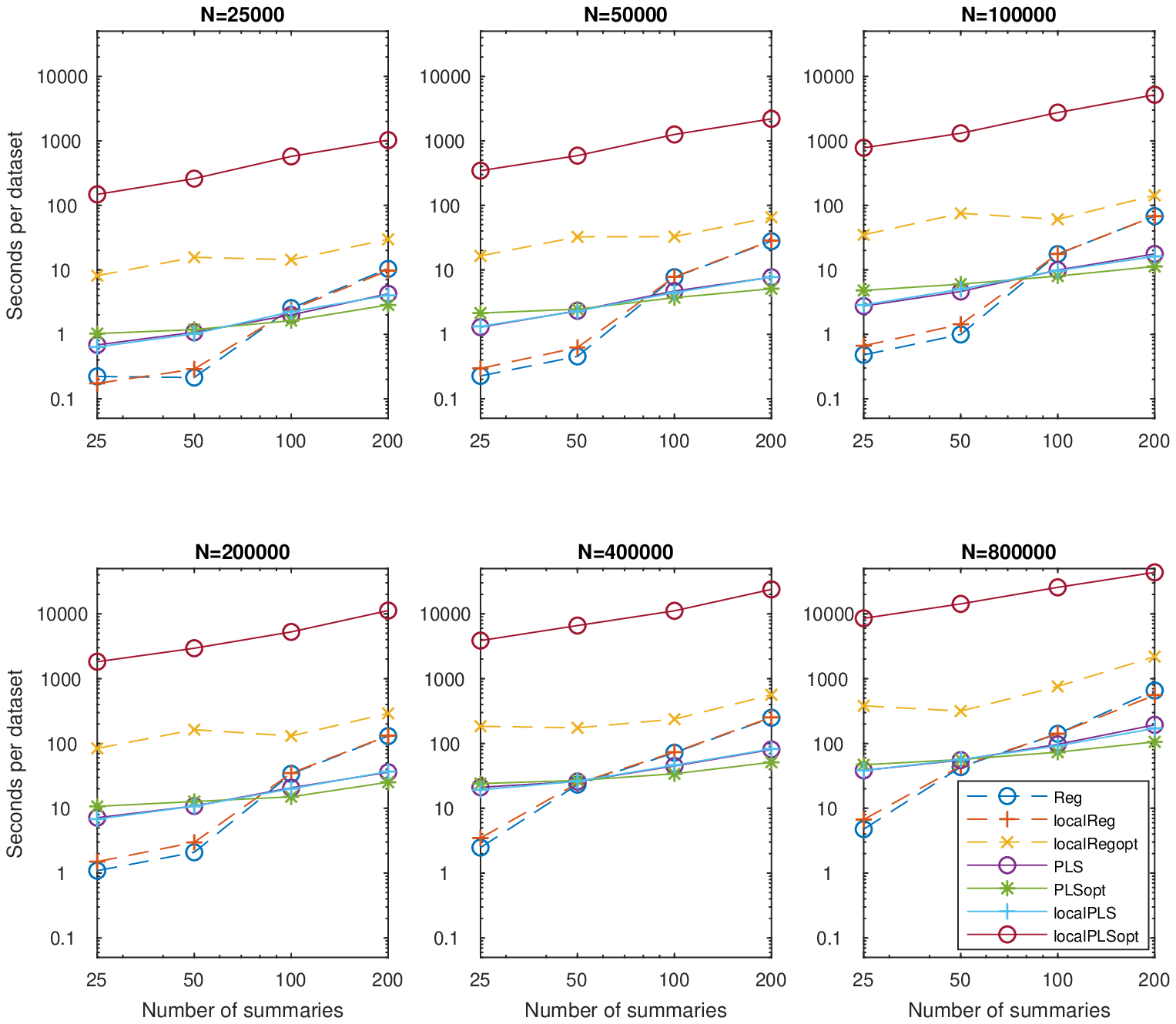}
	\captionof{figure}{Average runtime of different dimension reduction techniques with the $g$-and-$k$-distribution for different numbers of simulations and candidate summaries. Each panel shows the average runtime per dataset for number of simulations as indicated in the title. Each line shows the runtime for one method as a function of the number of summaries. Both the number of summaries and the runtime are shown on logarithmic scale.}
	\label{fig:gkruntime}
\end{center}